# Rewritable Chirality of Metasurfaces with Permittivity-Asymmetric Flatband Quasi-Bound States in the Continuum

*Tao Jiang, Haiyang Hu, Dmytro Gryb, Leonardo de S. Menezes, Maxim V. Gorkunov, Alexander A. Antonov*, and Andreas Tittl**

T. Jiang, A. Tittl
Institute of Photonics, Hamburg University of Technology, 21073 Hamburg, Germany
E-mail: andreas.tittl@tuhh.de

T. Jiang, H. Hu, D. Gryb, L. de S. Menezes, A. A. Antonov, A. Tittl
Chair in Hybrid Nanosystems, Nanoinstitute Munich, Faculty of Physics, Ludwig-Maximilians-Universität München, 80539 München, Germany
E-mail: a.antonov@physik.uni-muenchen.de

L. de S. Menezes
Departamento de Física, Universidade Federal de Pernambuco, 50670-901 Recife-PE, Brazil

M. V. Gorkunov
Theoretical Physics and Quantum Technologies Department, National University of Science and Technology 'MISIS', Moscow, Russia
National Research Nuclear University 'MEPhI' (Moscow Engineering Physics Institute), Moscow, Russia

Funding:
This project was funded by the Deutsche Forschungsgemeinschaft (DFG, German Research Foundation) under grant numbers EXC 2089/1–390776260 (Germany's Excellence Strategy) and TI 1063/1 (Emmy Noether Program), the Bavarian program Solar Energies Go Hybrid (SolTech), and Enabling Quantum Communication and Imaging Applications (EQAP), and the Center for NanoScience (CeNS) at LMU. It is also funded by the European Union (ERC, METANEXT, 101078018 and EIC, OMICSENS, 101129734). Views and opinions expressed are, however, those of the author(s) only and do not necessarily reflect those of the European Union, the European Research Council Executive Agency, or the European Innovation Council and SMEs Executive Agency (EISMEA). Neither the European Union nor the granting authority can be held responsible for them. M.G. acknowledges the support by the Federal Academic Leadership Program Priority 2030 (NUST MISIS Strategic Technology Project "Quantum Internet")

Keywords: bound states in the continuum, optical metasurfaces, flatband, chirality, chirality encoding

## Abstract

Flatband eigenstates are widely applied to enhance angle-robust light-matter interactions in metaphotonics. However, controlling the polarization of flatbands remains challenging, as it is usually fixed once the metasurface is fabricated, with no options of post-fabrication modification. Here, we present a rewritable permittivity-asymmetric quasi-bound state in the continuum (ε-qBIC) metasurface platform, where selective polymethyl methacrylate (PMMA) coating of a silicon double-nanorod unit cell establishes a circularly polarized flatband state. By varying the PMMA thickness, the polarization of this state can be further controlled in the range from the right-circular to linear and to left-elliptical. The flatbands maintain stable resonance positions and robust far-field polarizations for the incidence angles up to 10 degrees. Importantly, the PMMA layer can be removed, recoated, and re-patterned on the same nanostructure, providing a pathway to rewrite the optical response. Building on this capability, we experimentally demonstrate the chirality encoding by spatially selective PMMA coating. The results establish a practical strategy for realizing high quality factor flatband metasurfaces with rewritable chirality, thus opening opportunities for applications in chiral encoding and chiroptical photonic devices in general.

*Tao Jiang, Haiyang Hu, and Dmytro Gryb contributed equally to this work.*

## 1. Introduction

Metasurfaces provide a powerful platform for manipulating light at the subwavelength scale, enabling precise control over amplitude, phase, and polarization through engineered nanostructures[1–3]. This capability is further enhanced in all-dielectric metasurfaces through bound states in the continuum (BICs), which are nonradiative modes embedded in the radiation continuum[4,5]. Among them, symmetry-protected BICs which non-radiative character is granted by the point symmetry of the metasurface, offer convenient ways to engineer the coupling to the far-field by weak symmetry-breaking perturbations. As they are transformed into quasi-BICs (qBICs), their quality factors (Q-factors) and polarization are precisely controlled by the degree of structure asymmetry[6]. Owing to their strong field confinement and such qBIC metasurfaces offer broad opportunities for the manipulation of light-matter interactions, including dispersion manipulation [7] and far-field polarization control[8] across linear, elliptical and circular polarization states associated with optical chirality[9].

Optical chirality represents one of the most intriguing and fundamentally rich phenomena in photonics[10–13]. Optically chiral metasurfaces interact differently with right- (RCP) and left- (LCP) circularly polarized light and are desired for applications, such as chirality encoding[14], enantiomers detection[15], chiral emission[16,17] and photodetection[18]. Of particular interest is maximum chirality[9,19], in which the object is transparent to incident light of one circular polarization while resonantly interacts with the opposite polarization. The crucial step to achieve maximum chirality using qBIC metasurfaces is to completely break all the mirror symmetries of the structure, including the out-of-plane symmetry[20–23]. Although the presence of a substrate introduces out-of-plane symmetry breaking, substrate-induced chirality remains mostly as a theoretical concept[24,25]. In practice, advanced nanofabrication techniques have been explored to introduce out-of-plane symmetry breaking, including multistep lithography[26], grayscale lithography[27], and slanted reactive ion etching[16,28]. Most of the intrinsically chiral metasurfaces are designed to support circularly polarized eigenstates coupled to normally incident light, i.e., at the Γ-point in momentum space.

Extending chiral responses over a broader momentum space poses a significant challenge, as both the resonance spectral position and the far-field polarization eigenstates strongly depend on the in-plane wavevector. Flatband metasurfaces, characterized by nearly vanishing group velocity, provide a unique regime in metaphotonics in which resonances become dispersionless[29,30]. The energy of optical states is nearly independent of momentum, and their high quality factor (high-Q) provides strong localization of electromagnetic fields that enhances many useful optical phenomena such as strong coupling[31,32] and slow-light effects[33]. Flatbands can be realized by the monoclinic or Kagome lattice[34,35] topology-driven mechanisms[32], extreme subwavelength structuring[36], or multilayer coupling schemes[37,38]. Furthermore, a platform that combines flatbands with chirality is highly desirable for applications such as quantum imaging[39,40], monolithic photodetectors[41], enhanced emission[42], and flatband lasing[43]. Experimental demonstrations of chiral flatband metasurfaces have been realized either by adjusting the coupling strength between propagating modes, with operation limited to incident angles of 5° [44], or through nontrivial monoclinic lattice design[35]. These approaches rely on structural geometries that are fixed once fabricated, thereby restricting the post-fabrication tunability of the chiral response. Changing the metasurface response therefore requires the incorporation of additional mechanisms, such as phase-change materials[45], micro-electro-mechanical systems[46], or kirigami-based[47] platform. However, the optical response remains vulnerable to structural deviations, leaving minimal fabrication tolerance.

Here, we experimentally demonstrate a rewritable metasurface platform with flatbands and maximally intrinsic chiral permittivity-driven quasi-BICs (ε-qBICs). By partial cladding of a metasurface with a polymethyl methacrylate (PMMA) layer, one can induce controlled perturbations that affect near-field mode profiles and allow the mode to couple into radiation channels, forming an ε-qBIC[48–51]. Building on this concept, we extend the platform toward chirality control, where the far-field polarization can be continuously tailored from ideal circular to linear and further to elliptical polarization with opposite handedness by varying the PMMA thickness. All resulting polarization states preserve flatband characteristics, highlighting the potential for flexible chiral flatband functionalities. Moreover, the rewritable feature of the PMMA cladding relaxes fabrication constraints as the optical response can be repeatedly corrected, and optimized after fabrication. To illustrate the rewritability, we remove, recoat, and re-pattern the PMMA layer on the same silicon metasurface and further demonstrate chirality encoding. These findings establish the rewritable chiral ε-qBIC platform as a promising candidate for advanced applications, including chiral encryption, strong coupling and polarization detection.

## 2. Results and discussion

### 2.1 Concept of rewritable chirality of flatbands in the ε-qBIC metasurface

The original metasurface unit cell consists of two silicon nanorods with identical lengths and heights, but different widths and diverged by an opening angle $\alpha$. Due to the presence of the out of-plane symmetry $\sigma$, the system remains achiral, and the underlying antiparallel-dipole-like qBIC couples identically to RCP and LCP light, as evidenced by $\Delta T = T_{LL} - T_{RR} = 0$ under normal incident (Figure 1a). Although the presence of the substrate breaks the out-of-plane symmetry $\sigma$ ($xy$-plane), its impact on the qBIC optical chirality is negligible (Figure S1). Therefore, we assume the existence of $\sigma$ mirror symmetry in the following discussion. In previous studies[22,24], the out-of-plane mirror symmetry was broken by placing identical rectangular nanorods on different faces. In contrast, here we cover the thinner nanorod by a dielectric medium, achieving a similar symmetry breaking.

In this work, we utilize a PMMA layer with refractive index $n_{cov}$ of 1.5 and thickness $h$ as a cladding material (Figure 1b). It is worth mentioning that our platform also allows to incorporate various cladding materials, such as porous $SiO_2$[52] ($n_{cov}$ = 1.2), or $Al_2O_3$[53] ($n_{cov}$ = 1.8) (Figure S2). This selective encapsulation modifies the local environmental permittivity within the unit cell, enabling the chiral ε-qBIC with $\Delta T$ = 1.0. This indicates the co-polarized component $T_{LL}$ is fully decoupled while the $T_{RR}$ remains highly modulated (Figure S2), highlighting the coupling of the chiral ε-qBIC exclusively to the RCP light. The cross-polarized terms $T_{RL}$ and $T_{LR}$ are negligible and under lossless conditions, RCP light is reflected without a change in handedness (Figure S3), consistent with the theory[21,54]. Through an etching process, such as reactive ion etching (RIE), different PMMA thicknesses are achieved, leading to an evolution of chirality responses from right-circular to right-elliptical, linear, and finally left-elliptical polarization states, as shown in Figure 1c. To demonstrate the rewritability, the PMMA layer is removed through a cleaning process, such as RIE or wet etching using acetone, recovering the pristine silicon metasurface. By recoating and rewriting the PMMA layer, the polarization states of chiral ε-qBICs can be repeatedly corrected or redesigned.

Remarkably, the chiral ε-qBIC metasurfaces with PMMA cladding of varying thickness $h$ exhibit pronounced flatbands along the $k_y$-direction across different polarization states (Figure 1d-f), as exemplified by the right-circular and left-elliptical cases associated with different PMMA thicknesses in Figure 1c. This stability against the incidence angle can be understood as follows: the qBIC is primarily attributed to antiparallel electric dipoles, and, as experimentally demonstrated by scattering scanning near-field optical microscopy, the coupling between neighboring unit cells is significantly stronger along the $x$-direction than along the $y$-direction[55]. Obliquely incidence light excites dipoles with a phase shift, and depending on the inter-unit-cell coupling, the resonance wavelength exhibits either strong dispersion along the $k_x$-direction or a flatband response along the $k_y$-direction, as detailed in Figure 2d.

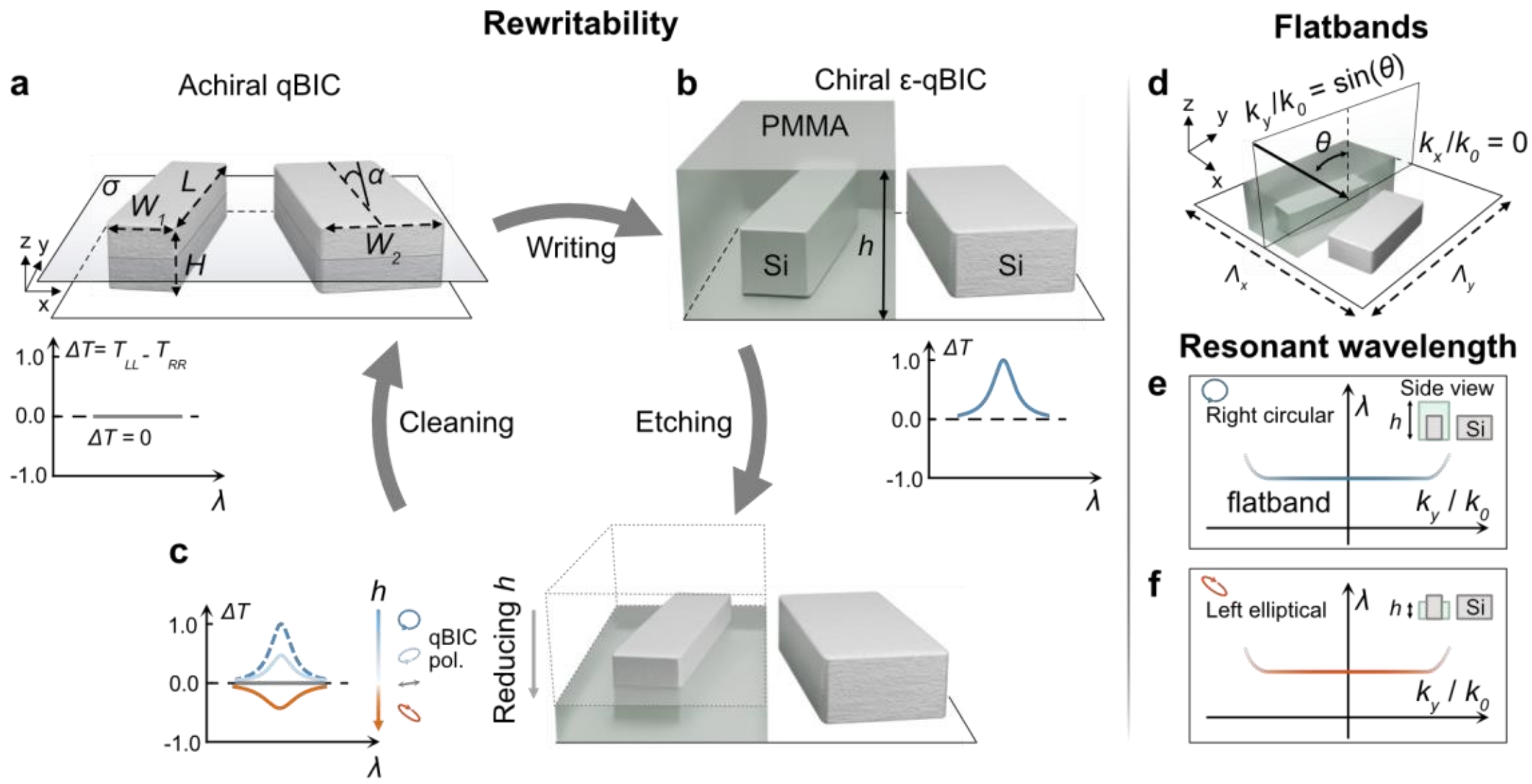

**Figure 1. Concept of rewritable chirality of a flatband ε-qBIC metasurface. a** An achiral qBIC silicon metasurface with out-of-plane mirror symmetry $\sigma$. The inset shows transmittance difference $\Delta T = T_{LL} - T_{RR} = 0$ under normal incidence. $T_{fi}$ denotes the transmittance, where $i$ and $f$ correspond to the input polarization and the analyzed polarization, respectively. **b** Breaking out-of-plane mirror symmetry $\sigma$ to achieve a maximally chiral ε-qBIC with $\Delta T = 1$ by writing a PMMA pattern of thickness $h$ on one of silicon nanorods. **c** Different heights of PMMA are achieved through an etching process. The inset indicates the evolution of $\Delta T$ and the corresponding polarization states as the PMMA thickness $h$ varies. A cleaning process removes the PMMA, followed by recoating, yielding rewritable ε-qBIC chirality and allowing repeated correction or redesign of optical chirality in a cyclic manner. **d** Schematic of the structure under illumination at an incidence angle $\theta$ in the $k_x / k_0 = 0$ plane, where $k_0$ is the free-space wavenumber. **e** and **f** Resonant wavelength of the structure along the $k_y$-direction for two different PMMA thicknesses $h$, exhibiting flatband dispersion.

## 2.2 Theoretical analysis of ε-qBIC Flatbands

Our PMMA coverage strategy provides flexible control over the chirality of ε-qBIC metasurfaces through the PMMA height *h*, while the underlying silicon nanorods remain structurally unchanged. We consider nanorods with length $L$ = 288 nm, height $H$ = 130 nm, widths $W_1$ = 65 nm and $W_2$ = 145 nm, and opening angle $\alpha$ = 7.5°. The nanorods are arranged in a square unit cell with periods $\Lambda_x = \Lambda_y$ = 500 nm and are placed on a $SiO_2$ substrate. For simplicity, refractive indices of silicon and $SiO_2$ are set to 3.83 and 1.45, respectively. In this configuration, as shown in Figure 2a, the ε-qBIC exhibits a near-maximum transmittance difference $\Delta T = T_{LL} - T_{RR}$ over a wide range of PMMA thickness *h*, which can even be extended to 270-600 nm (Figure S4). This corresponds to more than a 25-fold improvement in fabrication tolerance compared to the previously reported chiral structure composed of two nanorods with different heights[22]. For a metasurface with a PMMA thickness of 400 nm, the transmittance spectra $T_{LL}$ and $T_{RR}$ show that the chiral ε-qBIC is completely uncoupled from the LCP light, demonstrating maximal chirality (Figure 2b). When the PMMA thickness is reduced to $h$ = 200 nm, a distinct dip appears in the $T_{LL}$, while the $T_{RR}$ still remains low, indicating that the chiral ε-qBIC now exhibits right elliptical polarization. Further decreasing the PMMA thickness to $h$ = 150 nm and 100 nm induces a continuous transformation in the chiral response, where the polarization state evolves from right elliptical to linear and eventually to left elliptical, as also demonstrated in Figure 1c. To achieve left circular polarization, the opening angle $\alpha$ between two nanorods can be adjusted, providing a degree of freedom for tailoring the polarization state (Figure S5). However, this modification is limited to the design stage of silicon nanorods and does not allow post-fabrication tunability.

To study the evolution of the polarization state in more detail, Figure 2c shows the multipole decomposition of the near-fields of ε-qBIC with electric **P** (blue arrows) and magnetic 2/c·**M** (orange arrows) dipole moment contributions with both real (solid lines) and imaginary (dashed lines) parts (SI Note 1). First, we note that for all four polarization states, the electric dipole **P** contributes predominantly through its real part, whereas the magnetic dipole **M** contributes through its imaginary part, and the ratio between electric and magnetic dipoles amplitudes varies with the PMMA height. For the maximally chiral case ($h$ = 400 nm), the vectors representing the real part of **P** and the imaginary part of **M** have nearly equal magnitudes and are almost collinear, consistent with the condition of a perfect chiral emitter: $\mathbf{P} = -2i/c\cdot\mathbf{M}$. The extra doubling is due to nonzero electric quadrupole contribution[56] (SI Note 1). For the metasurface with $h$ = 200 nm, **M** still contributes mainly through its imaginary part and remains collinear with the real part of **P**. However, the absolute value of 2/c·**M** is nearly two times smaller than of the real part of **P**, resulting in pronounced right elliptical polarization of the ε-qBIC. For the metasurface with $h$ = 150 nm, only the real part of **P** contributes significantly to the eigenstate, while the contribution from **M** is negligible; therefore, the ε-qBIC becomes linearly polarized. Finally, the eigenstate of the metasurface with $h$ = 100 nm exhibits almost the same multipole composition as the case with $h$ = 200 nm, differing only in the sign of the imaginary part of **M**, which leads to left elliptical polarization of the ε-qBIC. Thus, variation of the PMMA thickness modifies the underlying ε-qBIC multipole composition, leading to different far-field polarizations, which can be experimentally accessed through post-fabrication process.

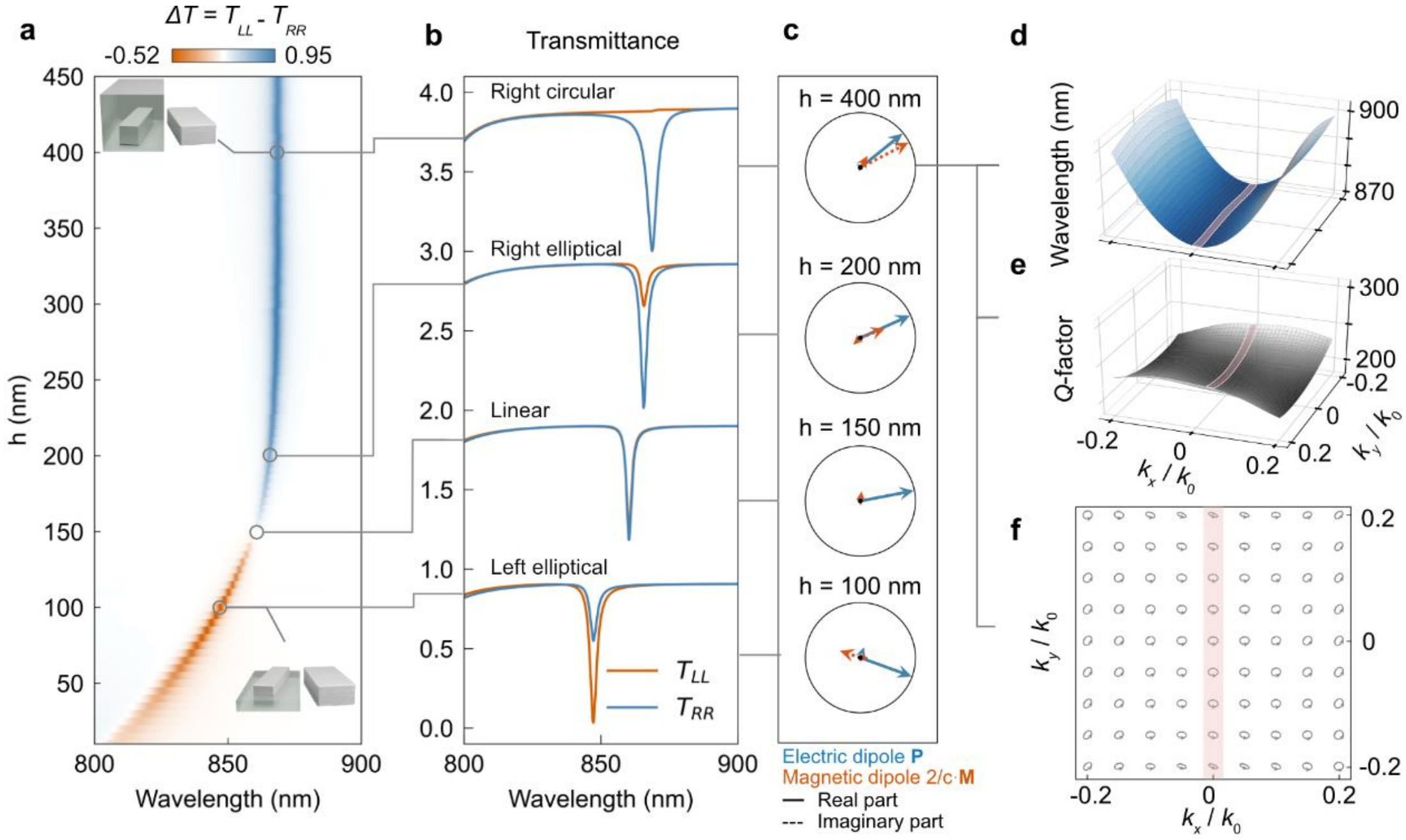

**Figure 2. Theoretical analysis of the chiral ε-qBIC metasurface. a.** Transmittance difference $\Delta T = T_{LL} - T_{RR}$ as a function of wavelength and PMMA thickness $h$. Insert: schematic of the unit cells with different PMMA heights. **b** Simulated transmittance spectra for four polarization states: right circular, right elliptical, linear, and left elliptical polarization. **c** Multipole decomposition analysis of the corresponding eigenstates. Blue and orange arrows indicate electric (**P**) and magnetic (2/c·**M**) dipole moment contributions, respectively, with their real and imaginary parts represented by solid and dashed lines. Simulated resonance position (**d**), Q-factor (**e**), and far-field polarization (**f**) in $k$-space for the maximally chiral ε-qBIC metasurface. The red shaded area highlights the flatband region.

To investigate the ε-qBIC in the $k$-space, Figure 2d presents simulated angular dispersion of the maximally chiral ε-qBIC metasurface with $h$ = 400 nm, where the resonance wavelength remains nearly unchanged over the range $-0.20 \leqslant k_y / k_0 \leqslant 0.20$. The dispersion has a half-pipe-shaped profile, featuring a pronounced flatband along the $k_y$-direction (highlighted in red). In the orthogonal $k_x$ direction, the resonance position shows a parabolic dispersion, enabling reversible and noninvasive tuning for modes, an essential feature for applications such as strong coupling[26]. Meanwhile, within the considered $k$-space range, the Q-factor varies moderately from 205 to 242 (Figure 2e), which is advantageous for experimental implementations requiring robust and angle-insensitive resonance characteristics. Importantly, the far-field polarization analysis demonstrates that the right circular polarization is maintained across a broad region of *k*-space along the $k_y$-direction in the range -0.15 $\leqslant k_y / k_0 \leqslant$ 0.15 (Figure 2f). This angular robustness of the maximal chiral response in terms of chirality, polarizations and Q-factors highlights the potential of the metasurface for reliable, large field-of-view applications.

### 2.3 Experimental Realization of Circularly Polarized ε-qBIC Flatbands

To experimentally confirm the ability of our platform to provide flatbands with various polarizations, we first focus on a metasurface supporting a circularly polarized ε-qBIC, which was fabricated by a two-step nanofabrication process (see Methods). First, silicon nanorods with the opening angle $\alpha$ = 6.5° were patterned onto a $SiO_2$ substrate using standard electron-beam lithography (EBL) followed by dry etching. In the second fabrication, a PMMA resist layer was spin-coated onto the pre-patterned metasurface and selectively exposed. After development, the PMMA layer selectively covered the defined region on the metasurface, introducing controlled out-of-plane asymmetry. The PMMA thickness $h$ was further optimized through RIE process, allowing precise adjustment over structural perturbation.

The combined asymmetry from both the in-plane opening angle $\alpha$ and the out-of-plane PMMA cladding thickness $h$ gives rise to a strongly chiral ε-qBIC, which is demonstrated experimentally by transmittance measurements under normal incidence in Figure 3a. All four components of the circular polarization transmittance ($T_{RR}$, $T_{LL}$, $T_{RL}$, $T_{LR}$), were measured to comprehensively evaluate the optical chirality of the ε-qBIC metasurface. A prominent resonance of $T_{RR}$ is observed at the wavelength of ~870 nm, while $T_{LL}$ remains unaffected, indicating that the ε-qBIC is in the maximally chiral regime and is decoupled from LCP. The resulting transmittance difference $\Delta T = T_{LL} - T_{RR}$ = 0.45 remains below the theoretical limit of unity, where $\Delta T$ is expected to approach the ultimate limit in the absence of measurement- and fabrication-induced imperfections[21,22], such as optical aberrations, geometric variations in the PMMA and silicon nanorods, and finite-size effects of the metasurface[57,58]. The polarization conversion components $T_{RL}$ and $T_{LR}$ remain negligible throughout the spectral window. Figure 3b presents an SEM image of the fabricated ε-qBIC metasurface with a PMMA cladding and silicon structural parameters including a period of $\Lambda_x = \Lambda_y$ = 460 nm, length $L$ = 310 nm, height $H$ = 117 nm, $W_1$ = 86 nm and $W_2$ = 178 nm. The fabricated PMMA thickness of $h$ = 290 nm satisfies the condition for maximum chirality (270 nm $\leq h \leq$ 600 nm), as demonstrated in Figure S4.

Next we measure the ε-qBIC response in *k*-space to investigate chiral flatbands by the experimental setups (see Methods, Figures S11 and S12). The *k*-space optical setup operates in reflectance mode, while transmittance are obtained from normal-incidence measurements using the chiral optical setup. The experimental co-polarized reflectance spectra, $R_{LL}$ (top panel) and $R_{RR}$ (bottom panel) for different $k_y/k_0$ values ($k_x$ = 0), show excellent agreement with simulations in terms of the dispersive spectral features (Figure 3c). A reduction in reflectance amplitude and lower Q-factors are nevertheless observed, consistent with the aforementioned practical limitations of fabrication and measurements. The corresponding reflectance difference is shown in Figure S6. Despite these experimental limitations, a pronounced flatband is observed in the $R_{RR}$ map, extending up to $k_y/k_0 \approx 0.18$, corresponding to an incidence angle of approximately 10°, while the $R_{LL}$ remains negligible, confirming angularly robust maximally chiral ε-qBIC (Figure 3c). Interestingly, along the orthogonal direction $k_x$ ($k_y$ = 0), ε-qBIC remains maximally chiral and exhibits a parabolic dispersion, allowing continuous spectral shifting via the incidence angle (Figure S7). This enables alignment of the chiral ε-qBIC with excitonic or molecular absorption features, facilitating applications such as chiral vibrational strong coupling (SI Note 11).

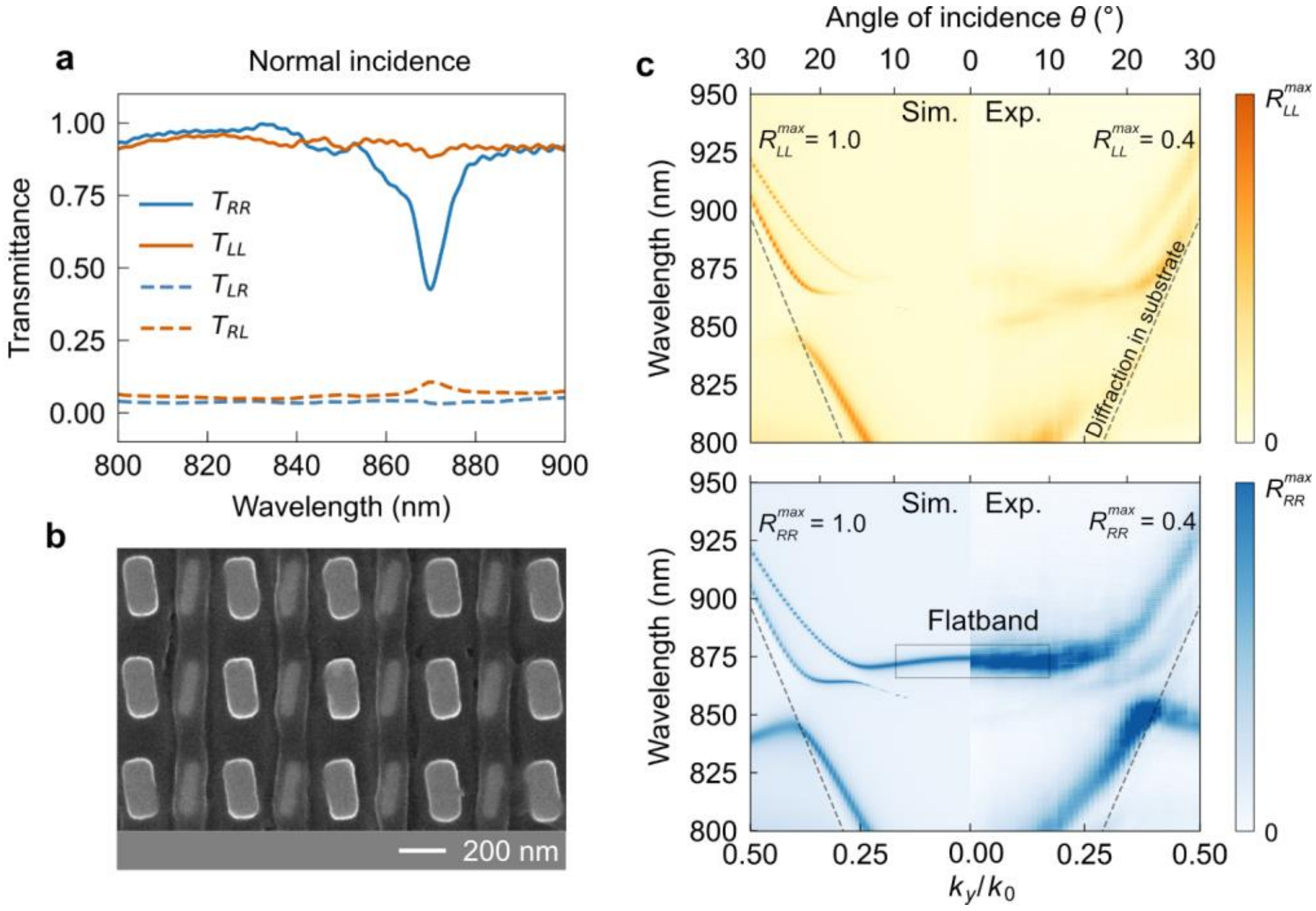

**Figure 3. Experimental characterization of a circularly polarized ε-qBIC flatband. a** Experimental transmittance spectra of the co-polarized components ($T_{RR}$ and $T_{LL}$) and cross-polarized components ($T_{LR}$ and $T_{RL}$) with maximally chiral ε-qBIC at the wavelength of ~870 nm under normal incidence. **b** SEM image of the fabricated ε-qBIC metasurface. Half of the silicon nanorods are covered by a PMMA cladding. **c** Simulated (left) and experimental (right) co-polarized reflectance $R_{LL}$ (top panel) and $R_{RR}$ (bottom panel), as functions of $k_y/k_0$, showing a chiral flatfand for $k_y/k_0 \leq 0.18$ highlighted by a rectangular box. The dashed lines indicate the diffraction cutoff associated with the $SiO_2$ substrate.

## 2.4 Experimental Realization of Flatbands with Different Polarizations

The PMMA cladding thickness effectively controls the ε-qBIC far-field polarization and can be precisely etched by a thickness variation $\Delta h$ using a RIE process (Figure 4a). Figures 4b and 4c present the experimental co-polarized reflectance spectra, $R_{LL}$ (left panel) and $R_{RR}$ (right panel), as a function of $k_y/k_0$ with $k_x/k_0 = 0$, and the co-polarized transmittance spectra $T_{LL}$ and $T_{RR}$ at the Γ-point, respectively. The corresponding reflectance spectra with parabolic dispersion along the orthogonal direction $k_x$ with $k_y/k_0 = 0$ are shown in Figure S8. First the system exhibits a right elliptical polarization state, where the ε-qBIC exhibits higher $R_{RR}$ compared to $R_{LL}$ (top panel of Figure 4b). This behavior is further confirmed by the transmittance spectra at Γ-point, where a small modulation of $T_{LL}$ indicates that the mode is no longer maximally chiral (Figure 4c). Next, following partial PMMA removal via the RIE process, the system achieves a linear polarization state, where $R_{RR}$ becomes comparable to $R_{LL}$ (middle panel of Figure 4b), while $T_{RR}$ has the same resonance modulation as $T_{LL}$ (Figure 4c). Meanwhile, the resonance wavelength shows a blueshift due to the reduced mode volume (Figure 4c), consistent with numerical simulations (Figure 2a). Upon further etching of the PMMA cladding, the eigenstate of the metasurface evolves to a left elliptical polarization state, where $R_{RR}$ becomes weaker than $R_{LL}$ (bottom panel of Figure 4b), and the modulations of the $T_{RR}$ and $T_{LL}$ are reversed compared to the initial right elliptical regime (Figure 4c).

All three regimes also exhibit dispersionless ε-qBICs in a broad range of $k_y/k_0$. To analyze these three flatbands, the ε-qBIC position from Figure 4b were extracted and fitted using a combination of linear and parabolic functions to obtain dispersion curves for each polarization state (Figure 4d). All three curves present a clear flatband feature extending up to $k_y/k_0 \approx 0.18$, in good agreement with the theoretical prediction (Figure S2d).

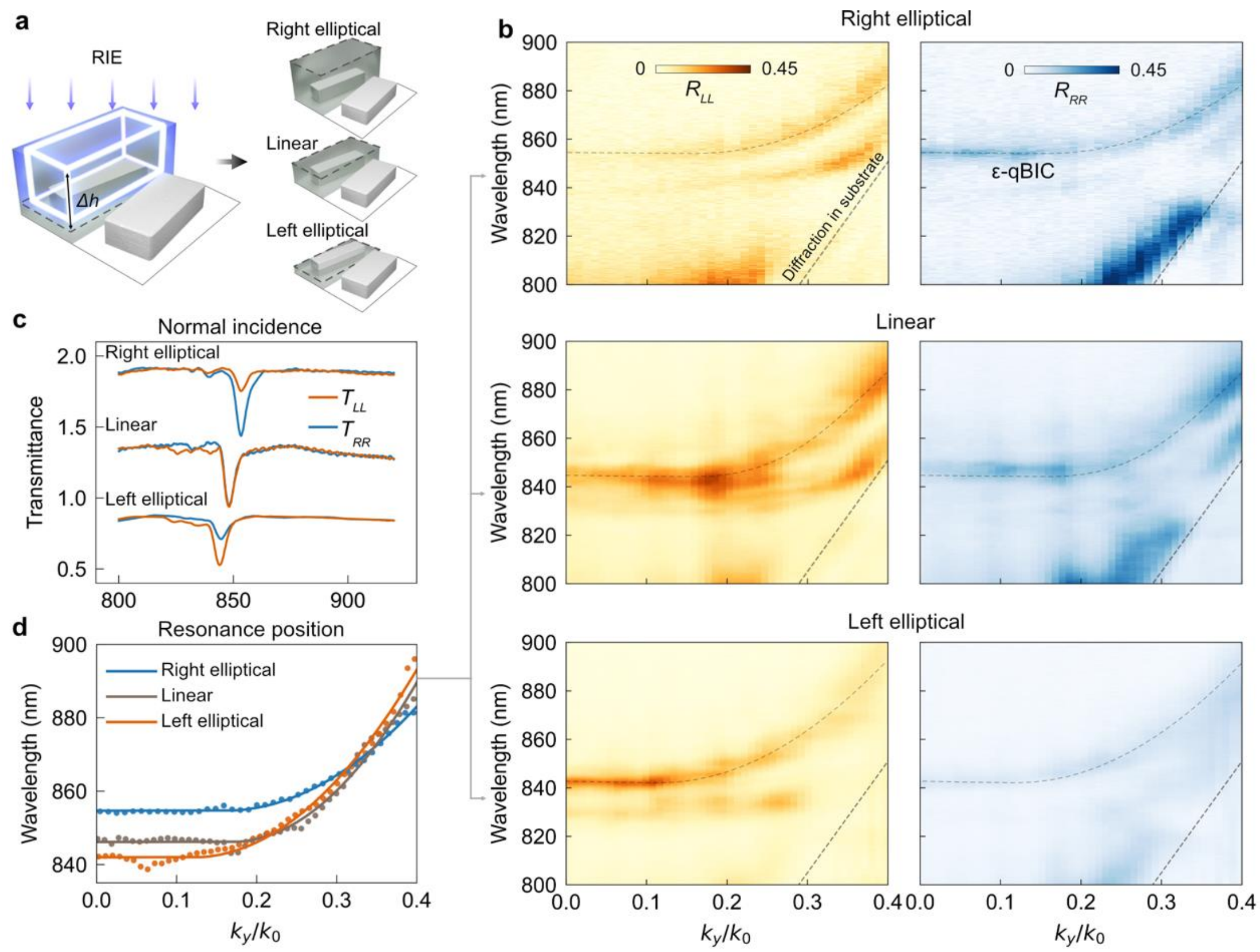

**Figure 4. Flatband polarization control of the ε-qBIC metasurface. a** Schematic of the RIE process used to reduce the PMMA layer by a thickness $\Delta h$ and, consequently, to access distinct ε-qBIC polarizations: right elliptical, linear, and left elliptical polarization. **b** Experimental co-polarized reflectance maps $R_{LL}$ (left panel) and $R_{RR}$ (right panel) as a function of $k_y/k_0$ with three polarization states. Dashed curves indicate the ε-qBIC positions. **c** Experimental co-polarized reflectance spectra under normal incidence for the corresponding polarization states. **d** Resonance position to each polarization state as a function of $k_y/k_0$. Dots represent maximum reflectance values corresponding to the ε-qBIC positions, and solid lines are the fitted curves. All three curves remain nearly dispersionless up to $k_y/k_0 \approx 0.18$.

## 2.5 Spatial Encoding of Chirality

Beyond flatbands with controlled chirality, our PMMA-based methodology enables a rewritable metasurface architecture with full control over the local cladding distribution, thereby modifying the local permittivity and optical signal. This allows PMMA cladding to be completely removed, recoated, and subsequently patterned and etched, enabling reprogramming of the polarization-selective response, which is not accessible in the conventional fabrication-fixed design, achieved through a two-step silicon deposition process[22]. Importantly, in contrast to the previously reported chiral information encoding scheme[14], where the encoded chirality is permanently defined after fabrication, our platform enables reversible reuse of the same silicon metasurface.

The chiral response can be rewritable by reintroducing PMMA, and further extended to spatially resolved chirality encoding, where local PMMA patterning defines position-dependent polarization states. Taken together, these capabilities provide full control over the cladding geometry, encompassing out-of-plane ($z$) thickness adjustment and in-plane ($x$-$y$) selectively spatial patterning.

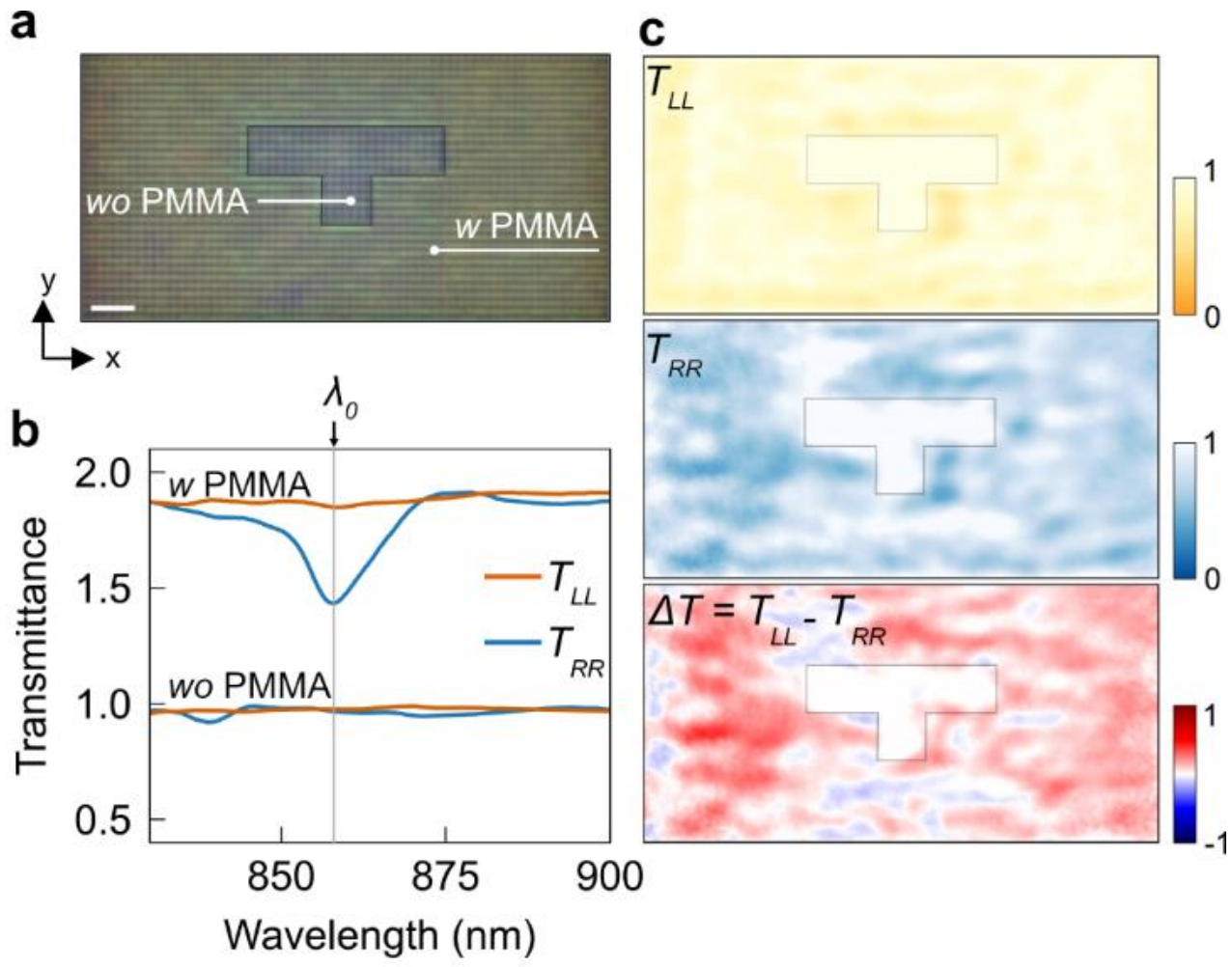

**Figure 5. Spatial encoding of chirality via selective PMMA patterning a** Optical microscope image of the metasurface, where the shape “T” region remains uncoated, while the surrounding area is covered with PMMA. The array size is 30 μm × 15 μm. Scale bar is 2.5 μm. **b** Corresponding co-polarized transmittance spectra for regions with and without PMMA. Region with PMMA cladding has maximally chiral ε-qBIC at the wavelength $\lambda_0$ = 860 nm. **c** Chiral imaging based on co-polarized transmittance $T_{RR}$, $T_{LL}$, and their difference $\Delta T$ at the wavelength of $\lambda_0$ = 860 nm. The dashed line and solid lines delineate the T-shaped region. In the $\Delta T$ image, the “T” region appears in white ($\Delta T \approx 0$), while the surrounding PMMA-covered area is predominantly in red ($\Delta T > 0$), indicating that the information is selectively encoded on the chiral ε-qBIC metasurface.

To experimentally demonstrate rewritable spatial encoding of chirality, PMMA was first removed and then spin-coated again onto the same silicon metasurface from previous sections, followed by EBL patterning of a letter “T”. The region inside the “T” remains uncoated with PMMA, while the surrounding area is covered by PMMA (Figure 5a). The corresponding co-polarized transmittance spectra for regions with and without PMMA are shown in Figure 5b. In the PMMA-coated region, the metasurface exhibits maximal chirality at the wavelength of $\lambda_0$ = 860 nm and couples only with RCP light. In contrast, the region without PMMA exhibits an achiral response at the same wavelength, with nearly identical transmittance for $T_{LL}$ and $T_{RR}$, and supports a regular linear polarized qBIC at the wavelength of ~810 nm, outside the spectral range shown in Fig. 5b.

This spatial contrast in chirality enables information encoding across the metasurface array, where the “T” region carries no chirality information ($\Delta T$ = 0) and the surrounding PMMA-covered area shows finite chirality ($\Delta T$ > 0). To verify this encoding mechanism, the array was illuminated with a narrow-linewidth tunable laser set at $\lambda_0$ = 860 nm, and the transmitted intensity images corresponding to $T_{LL}$ and $T_{RR}$ were recorded using a charge-coupled device (CCD) (Figure 5c). The $T_{LL}$ image shows relatively high intensity across the entire structure, since both the chiral and achiral regions hardly interact with LCP light. In contrast, the $T_{RR}$ image exhibits lower intensity in the PMMA-covered region outside the “T”, reflecting the right-handed chiral response of the metasurface. By calculating the transmittance difference, $\Delta T = T_{LL} - T_{RR}$, the encoded chirality distribution can be directly visualized. Most regions outside the “T” appear in red ($\Delta T$ > 0), whereas the region inside the “T” appears white ($\Delta T \approx 0$), clearly revealing the spatially encoded chiral pattern. Even though the “T” shape can be clearly discerned, the $\Delta T$ contrast outside the “T” region is not spatially uniform due to optical aberrations and proximity effects during PMMA patterning by EBL that alter the out-of-plane asymmetry and thus the local chirality of the ε-qBIC. The chirality-encoded image also remains observable at an incidence angle of 5° along the $k_y/k_0$ direction, confirming the robustness of the spatial encoding mechanism within the flatband regime (Figure S9). Due to the very shallow depth of field of the high-NA, high-magnification objective, imaging the rotated metasurface at larger rotation angles is severely constrained. Nevertheless, chirality stability at higher angles has previously been confirmed by both simulations and experiments, remaining nearly dispersionless for incident angles up to 10° (Figure 3c). These results demonstrate that PMMA-enabled permittivity control opens a new paradigm of rewritable chiral metasurfaces, where chirality is no longer a fixed structural property but a controllable degree of freedom. It enables potential photonic applications, including rewritable polarization metasurfaces, spatial polarization multiplexing, on-demand chiral optics for sensing and tailored light-matter interactions.

## 3. Conclusion

We have experimentally realized rewritable chiral flatband ε-qBIC metasurfaces with right-circular, right-elliptical, linear, and left-elliptical polarization states, by exploiting the environmental permittivity as a novel degree of freedom. This is achieved through selective PMMA coating of silicon nanorods of the same height. The coating provides the mirror symmetry breaking required for geometric chirality allowing us to avoid slanted or multistep lithography of silicon nanostructures. The resulting metasurfaces exhibit maximal chirality that is sustained across a broad range of PMMA thicknesses, representing more than a 25-fold increase in fabrication tolerance compared to the chiral structure composed of two nanorods with different heights[22]. Beyond this enhanced fabrication tolerance, adjusting the PMMA height enables the realization of different polarization states while maintaining robust resonance positions and far-field polarization across a wide range of momentum space. Notably, we observe flatband dispersion for incident angles up to approximately 10° along the $k_y/k_0$ direction ($k_x$ = 0) for all polarization states, highlighting the angular robustness and controllability of the platform. In addition, the ε-qBIC metasurface exhibits a parabolic dispersion along the orthogonal direction. This enables the resonance position control and facilitates applications in polaritonic systems, including strong coupling with excitonic, plasmonic, and vibrational resonances, without requiring actively tunable platforms.

Furthermore, the ability to remove, recoat, and re-pattern the PMMA layer enables a rewritable platform for chirality encoding across the metasurface. This rewritable capability is experimentally demonstrated by re-patterning the PMMA layer on the same metasurface, and the chirality encoding functionality is confirmed via chiral imaging based on the transmittance difference between the co-polarized components $T_{RR}$ and $T_{LL}$, where a "T"-shaped region without PMMA is surrounded by areas covered with the PMMA layer. Our platform enables full control over the cladding geometry and opens new possibilities for advanced chiral encryption and information encoding. Importantly, PMMA thickness can be controlled not only via RIE but also through grayscale EBL without changing the underlying nanorods, where different PMMA heights determined by the local electron dose induce corresponding changes in the chirality [59].

The chiral flatband ε-qBIC platform also provides a versatile framework for integrating functional materials. By incorporating emissive components, such as lasing dyes[16] or quantum dots[60], the metasurface can be extended toward applications in flatband lasing and strong light-matter coupling. The concept is not limited to PMMA and can be generalized to other material platforms, such as porous $SiO_2$ and $Al_2O_3$, enabling access to a wide range of refractive indices and thus diverse chiral responses while preserving flatband behavior. The platform can also be compatible with actively tunable materials, including phase-change materials[45] and liquid crystals[61], offering a route toward dynamically reconfigurable chiral photonic devices. We believe that our experimentally demonstrated chiral flatband ε-qBIC metasurface platform paves the way for diverse applications, establishing it as a compelling candidate for future chiral photonic technologies, including optical anti-counterfeiting, enantioselective detection, nonlinear optics, and rewritable chiral nanophotonic devices.

## 4. Methods

### 4.1 Numerical simulations

The optical response of the ε-qBIC metasurfaces were numerically investigated using the frequency domain solver in CST Studio Suite. Simulations were performed on a single unit cell of the metasurface under periodic boundary conditions in the *x*- and *y*-directions to represent an infinite array. The metasurface was excited from the top by left- and right- circularly polarized (LCP and RCP) plane waves to evaluate polarization-dependent transmittance properties. For silicon, $SiO_2$ and PMMA, constant refractive indices of 3.83, 1.45 and 1.50 were used, respectively, while material losses were neglected. Mesh settings were optimized to resolve subwavelength features and capture narrow ε-qBICs accurately.

Characteristics of the ε-qBIC in momentum space were numerically investigated using the Electromagnetic Waves Frequency Domain module of COMSOL Multiphysics in 3D mode using a previously developed approach[62]. The tetrahedral spatial mesh for finite element method was automatically generated by COMSOL's physics-controlled preset. Simulations were performed within a rectangular spatial domain containing a single metasurface unit cell with periodic boundary conditions applied to its sides

### 4.2 Nanofabrication

Fabrication of the chiral ε-qBIC metasurfaces was carried out through a two-step electron-beam lithography (EBL) process, incorporating both top-down silicon etching and selective PMMA patterning. A silicon layer was first deposited on $SiO_2$ substrates using plasma-enhanced chemical vapor deposition (PECVD, Oxford instruments). To enable precise alignment for subsequent fabrication steps, a set of gold alignment markers was patterned prior to metasurface fabrication. This was achieved by sputtering a 30 nm gold layer (Angstrom deposition system) onto the silicon, which was subsequently patterned to form the marker system.

For the silicon metasurfaces patterning, a layer of PMMA 950K A4 was spin-coated onto the substrates, followed by the deposition of a conductive ESpacer 300Z layer. Electron beam lithography (Raith eLINE Plus system) was applied for patterning the designed silicon metasurfaces with the assistance of pre-defined gold marker system. Development was performed in a 3:1 solution of isopropyl alcohol (IPA) and methyl isobutyl ketone (MIBK) for 135 s. A 50 nm chromium layer was then deposited to serve as a hard mask, followed by lift-off in Mircroposit Remover 1165 overnight at 80 $^{o}$C. Silicon etching was performed using a PlasmaPro 100 ICP-RIE system (Oxyford Instruments). After etching, the Cr hard mask was removed using a standard wet chromium etchant.

In the second EBL step, a layer of PMMA 950K A4 was spin-coated onto the fabricated silicon metasurfaces. Alignment was performed on the same gold marker system, and only PMMA in the defined region was exposed, followed by the same development procedures. The PMMA etching was carried out using mixture of $O_2$ and Ar plasma in a PlasmaPro 100 ICP-RIE system. To completely remove the PMMA from the silicon metasurface, the sample was first cleaned in acetone for 5 min, followed by $O_2$ plasma treatment for 5 min.

### 4.3 Chiral optical measurement

Chiral transmission spectra were measured using a custom-built transmission microscope setup on an optical table, illuminated by a supercontinuum white light laser (SuperK Extreme, NKT Photonics) operating at 5% power and a 0.302 MHz repetition rate (Figure S11). Linear polarization (horizontal or vertical) was generated using a broadband linear polarizer (LPVIS100, Thorlabs, 550-1500 nm), and converted to circular polarization using a broadband quarter-wave plate (QWP, RAC4.4.20, B-Halle, 500–900 nm). The QWP was positioned directly beneath a 10× microscope objective (Olympus PLN, NA = 0.25), which served as a condenser to minimize mirror-induced depolarization effects. The light was focused onto the sample and collected by a 60× objective (Nikon MRH08630, NA = 0.7). To ensure uniform illumination over the metasurface array, the beam was slightly defocused. An aperture was placed in the collection path to isolate light transmitted through the metasurface, thereby suppressing background signals from surrounding areas. Transmitted light was directed either to a CCD camera (QICAM Fast 1394 Digital Camera, 12-bit) for imaging the metasurface in transmittance, or to a multimode fiber (Thorlabs M15L05, core size: 105 μm, NA = 0.22) coupled into a grating spectrometer (Princeton Instruments, 300 g/mm grating, blaze angle of 750 nm) with a spectral resolution of 0.13 nm. Cross- and co-polarized transmission spectra were recorded by placing a chiral polarization analyzer after the sample, consisting of a QWP (AQWP05-580, Thorlabs, 350-850 nm) and a broadband linear polarizer.

### 4.4 Angle-resolved *k*-space optical characterization

Angle-resolved *k*-space measurements were performed using a custom-built optical setup in a back-reflection configuration (Figure S12). A halogen lamp (Thorlabs SLS201L) provided broadband illumination, which was collimated and directed toward the metasurface via a 50:50 beamsplitter and a 60× objective (NA = 0.95). The same objective was used to collect the reflected signal. The back focal plane of the objective, corresponding to the *k*-space distribution, was imaged by a Fourier lens and relayed onto the entrance slit of a spectrometer through a 4f system formed with an additional lens. Polarization control was achieved using a linear polarizer (Thorlabs LPVISC100-MP2) followed by a quarter-wave plate (Thorlabs λ/4, SAQWP05M-700) oriented at 45°, generating circularly polarized incident light. Upon reflection, the beam passed again through the quarter-wave plate, converting it back to linear polarization. A second linear polarizer served as an analyzer, enabling polarization-resolved angle-dependent measurements. The signal was dispersed by a spectrometer (Princeton Instruments, 150 g/mm, blaze angle of 800 nm) and subsequently recorded by a CCD detector.

**Acknowledgements**

This project was funded by the Deutsche Forschungsgemeinschaft (DFG, German Research Foundation) under grant numbers EXC 2089/1–390776260 (Germany's Excellence Strategy) and TI 1063/1 (Emmy Noether Program), the Bavarian program Solar Energies Go Hybrid (SolTech), and Enabling Quantum Communication and Imaging Applications (EQAP), and the Center for NanoScience (CeNS). It is also funded by the European Union (ERC, METANEXT, 101078018 and EIC, OMICSENS, 101129734). Views and opinions expressed are however those of the author(s) only and do not necessarily reflect those of the European Union, the European Research Council Executive Agency, or the European Innovation Council and SMEs Executive Agency (EISMEA). Neither the European Union nor the granting authority can be held responsible for them. The work of M.V.G. was carried out within the State assignment of NRC "Kurchatov Institute".

**Data Availability Statement**

All data are available in the main text or the supplementary materials.

**Author contributions**

T.J., H.H., and D.G. contributed equally to this work and agree that the order of their names may be exchanged as useful to highlight their contributions in individual professional pursuits. T.J., H.H., D.G., A.A.A., and A.T. contributed to the conceptualization, data collection, and writing of the manuscript, developed the methodology and performed the data analysis. L.S.M., M.V.G., A.A.A., and A.T. contributed to the review and editing of the manuscript. L.S.M., M.V.G., A.A.A., and A.T. supervised the project. All authors reviewed and approved the final manuscript and agree to be accountable for all aspects of the work, ensuring its accuracy and integrity.

**Conflict of interest**

The authors declare that they have no conflict of interest.

## Supporting information:

# Rewritable Chirality of Metasurfaces with Permittivity-Asymmetric Flatband Quasi-Bound States in the Continuum

*Tao Jiang, Haiyang Hu, Dmytro Gryb, Leonardo de S. Menezes, Maxim V. Gorkunov, Alexander A. Antonov*, and Andreas Tittl**

T. Jiang, A. Tittl
Institute of Photonics, Hamburg University of Technology, 21073 Hamburg, Germany
E-mail: andreas.tittl@tuhh.de

T. Jiang, H. Hu, D. Gryb, L. de S. Menezes, A. A. Antonov, A. Tittl
Chair in Hybrid Nanosystems, Nanoinstitute Munich, Faculty of Physics, Ludwig-Maximilians-Universität München, 80539 München, Germany
E-mail: a.antonov@physik.uni-muenchen.de

L. de S. Menezes
Departamento de Física, Universidade Federal de Pernambuco, 50670-901 Recife-PE, Brazil

M. V. Gorkunov
Theoretical Physics and Quantum Technologies Department, National University of Science and Technology 'MISIS', Moscow, Russia
National Research Nuclear University 'MEPhI' (Moscow Engineering Physics Institute), Moscow, Russia

*Tao Jiang, Haiyang Hu, and Dmytro Gryb contributed equally to this work.*

## Contents

## 1 Multipole decomposition analysis

To address how the thickness of the PMMA cladding affects the polarization of ε-qBICs, we employ a previously developed methodology[1], which is briefly outlined in this section. The coupling coefficient between an eigenstate of a metasurface and a normally incident plane wave propagating along the $z$-direction, characterized by wavenumber k = ω/c and polarization unit vector **e**, can be given by an overlap integral:

$$m_{\mathrm{e}} \propto \int_V \mathbf{J}(\mathbf{r}) \cdot \mathbf{e} e^{i\mathbf{k}\cdot\mathbf{r}} dV$$

$$\approx \int_V \mathbf{J}(\mathbf{r}) \cdot \mathbf{e}(1 + i\mathbf{k}\cdot\mathbf{r}) dV$$

$$\approx -i\omega\mathbf{P}\cdot\mathbf{e} - i(\mathbf{k}\times\mathbf{M})\cdot\mathbf{e} + \frac{\omega}{6}(e_\alpha)(k_\beta)Q_{\alpha\beta} \quad \text{(S1)}$$

where $\mathbf{J}(\mathbf{r})$ is an eigenstate displacement current density of the metasurface, and the integration is performed over the volume of the silicon nanorods and the PMMA cladding.

The electric and magnetic dipole moments, along with the electric quadrupole moment, are introduced according to the standard formalism[2]:

$$\mathbf{P} = \frac{i}{\omega}\int_V \mathbf{J}(\mathbf{r}) dV \quad \text{(S2)}$$

$$\mathbf{M} = \frac{1}{2}\int_V \mathbf{r}\times\mathbf{J}(\mathbf{r}) dV \quad \text{(S3)}$$

$$Q_{\alpha\beta} = \frac{3i}{\omega}\int_V \left[r_\alpha J_\beta(\mathbf{r}) + r_\beta J_\alpha(\mathbf{r}) - \frac{2}{3}\delta_{\alpha\beta}\mathbf{r}\cdot\mathbf{J}(\mathbf{r})\right] dV \quad \text{(S4)}$$

Consistently with our experimental results, we consider circularly polarized incident waves with unit vectors $\mathbf{e}_\pm = \frac{1}{\sqrt{2}}(\mathbf{e}_x \pm i\mathbf{e}_y)$, which reduces Eq. (S1) to:

$$m_\pm \propto [\mathbf{P}_x \pm i\mathbf{P}_y] \pm \frac{i}{c}[\mathbf{M}_x \pm i\mathbf{M}_y] + \frac{i\omega}{6c}[Q_{xz} \pm iQ_{yz}] \quad \text{(S5)}$$

Since the qBIC current $\mathbf{J}(\mathbf{r})$ is mainly confined to the $xy$-plane ($Jz \approx 0$), one can simplify the multipole terms in Eq. (S5) and obtain the following expression for the coupling coefficient:

$$m_\pm \propto [\mathbf{P}_x \pm i\mathbf{P}_y] \pm \frac{2i}{c}[\mathbf{M}_x \pm i\mathbf{M}_y]. \quad \text{(S6)}$$

According to the Eq.(S6) the condition for maximum chirality $m_\pm = 0$ is satisfied when:

$$\mathbf{M} = \pm\frac{ic}{2}\mathbf{P} \quad \text{(S7)}$$

As previously noted in [1], (Eq.S7) differs from the conventional relation $\mathbf{M} = \pm ic\mathbf{P}$, commonly used to describe chiral molecular scattering[3] and emission[4], by an additional factor of 1/2. This factor originates from the nonvanishing contribution of the electric quadrupole $Q_{\alpha\beta}$, which remains nonzero due to the fixed orientation of the metasurface, unlike freely rotating molecules where orientational averaging suppresses the quadrupole contribution.

We employ the Eigenstate Solver of COMSOL Multiphysics to analyze the qBIC near fields of the metasurface shown in Figure 2 of the main text. Since the eigenstate is determined up to an unknown phase $\phi$, we choose it such that $\mathrm{Re}(\mathbf{P}e^{i\phi}) \cdot \mathrm{Im}(\mathbf{P}e^{i\phi}) = 0$, and plot the real and imaginary parts of $\mathbf{P}e^{i\phi}$ in Fig.2c.

Next we note that the magnetic dipole $\mathbf{M}$ depends on the choice of coordinate system. Since the expansion in Eq. (S1) was performed near $z = 0$, we slightly shift the coordinate origin by $\mathbf{r}_0 = (0,0,z_0)^T$, with $z_0 = 80$ nm chosen to approximately align the cross-sectional plane with the region of maximum field enhancement inside silicon nanorods. The resulting magnetic dipole components are then given by:

$$M_x = -\frac{e^{i\phi}}{2}\left[\int_V zJ_y(\mathbf{r})dV - z_0 \int_V J_y(\mathbf{r})dV\right] \quad \text{(S8)}$$

$$M_y = \frac{e^{i\phi}}{2}\left[\int_V zJ_x(\mathbf{r})dV - z_0 \int_V J_x(\mathbf{r})dV\right]. \quad \text{(S9)}$$

Accordingly, the real and imaginary parts of $2c^{-1}\mathbf{M}e^{i\phi}$ are plotted in Figure 2c.

## 2 Effect of the substrate on the optical chiral response

To study the effect of the substrate on the chiral response, we simulated the structure with a $SiO_2$ substrate. For the structure with $\sigma_1$ symmetry, it exhibits a linearly polarized qBIC (top panel of Figure S1a), with no chiral signal evident, as the transmittance difference remains zero (top panel of Figure S1b). Introducing an opening angle of $\alpha = 7.5°$ to each nanorod breaks the in-plane symmetry and results in a weak chiral response, with transmittance difference below 0.1 (bottom panel of Figure S1b). These results indicate that, although the presence of the substrate breaks the out-of-plane symmetry (*xy*-plane), its impact on the optical chirality of the qBIC remains negligible. It highlights that a pronounced chiral response requires an additional asymmetry in the local dielectric environment, such as selectively embedding one of the nanorods using a dielectric cladding layer (Figure 1b).

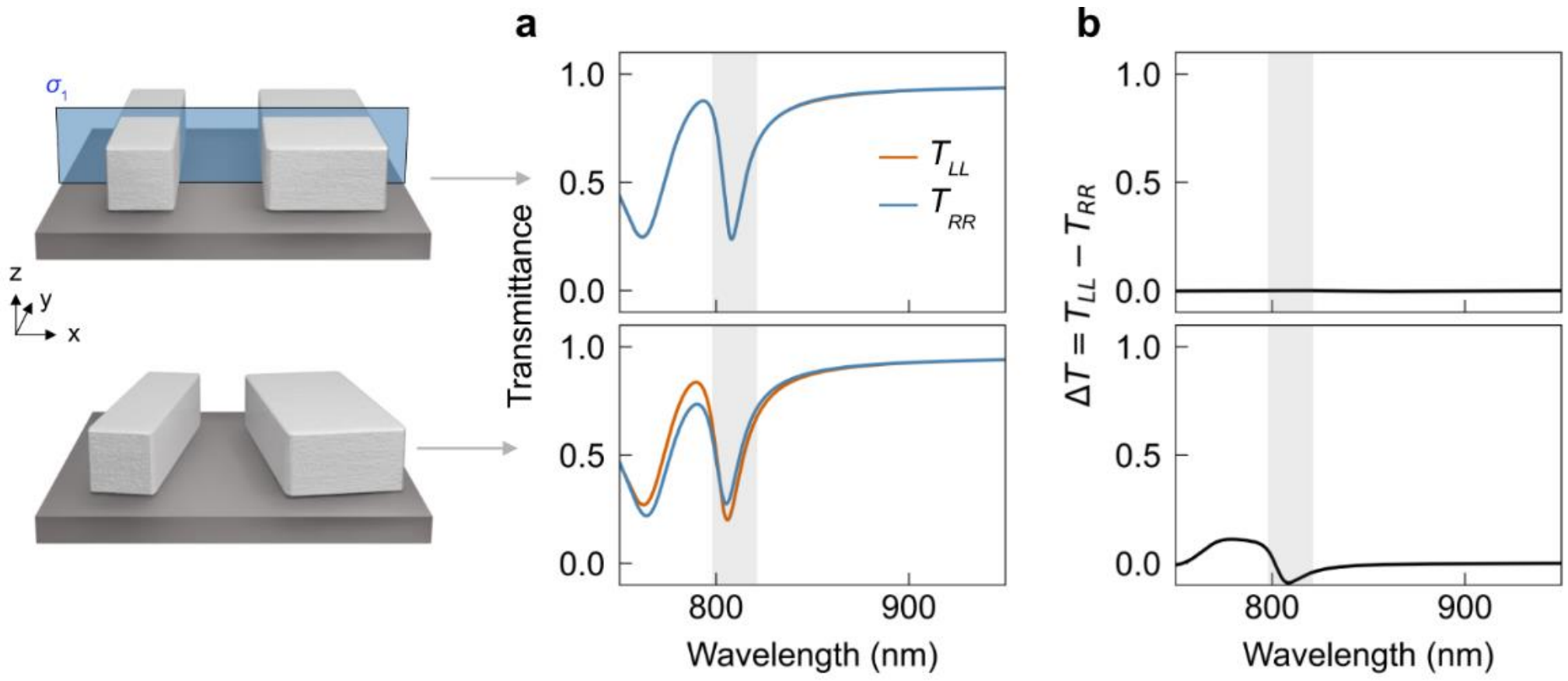

**Figure S1. Effect of the $SiO_2$ substrate on qBICs. a** Simulated transmittance spectra for two structures. The grey shaded regions indicate the qBIC positions, located at wavelengths of 808 nm (top panel) and 806 nm (bottom panel). **b** Corresponding transmittance difference $\Delta T = T_{LL} - T_{RR}$ as a function of wavelength.

## 3 Polarization of ε-qBIC as a function of cladding

The metasurface supports intrinsically chiral ε-qBIC and its optical response relies on both the refractive index ($n_{cov}$) and thickness ($h$) of the cladding. To understand how the cladding affects the resonance, we plot the far-field polarization map of the ε-qBIC as a function of $n_{cov}$ and $h$ in Figure S2. Our platform allows to incorporate various overlayer materials, such as porous $SiO_2$[5] ($n_{cov}$ = 1.2), PMMA[6] ($n_{cov}$ = 1.5) or $Al_2O_3$[7] ($n_{cov}$ = 1.8), thereby providing access to the corresponding polarizations of the ε-qBIC (Figure S2c). The shaded gray region highlights the PMMA refractive index used in this work. The size of each ellipse is inverse proportional to the Q-factor of the corresponding ε-qBIC where smaller ellipses represent narrower spectral linewidths and thus higher Q-factors. For $n_{cov}$ = 1.5, as investigated in this work, decreasing $h$ first leads to increasing of the Q-factor, as evidenced by the shrinking ellipse size. During this process, the far-field eigenstate evolves from right circular polarization ($h$ = 400 nm), through right elliptical polarization ($h$ = 200 nm), to linear polarization state ($h$ = 150 nm). At the certain $h$ value, the resonance approaches the condition of a restored symmetry-protected (accidental) BIC, as was demonstrated previously[8], resulting in a relatively high Q-factor. As $h$ is further reduced, the far-field polarization ellipse of the ε-qBIC reverses its handedness, while the Q-factor decreases. Furthermore, varying the PMMA cladding thickness $h$ preserves pronounced flatband behavior along the $k_y$ direction together with stable far-field polarization states in chiral ε-qBIC metasurfaces (Figure S2c, d).

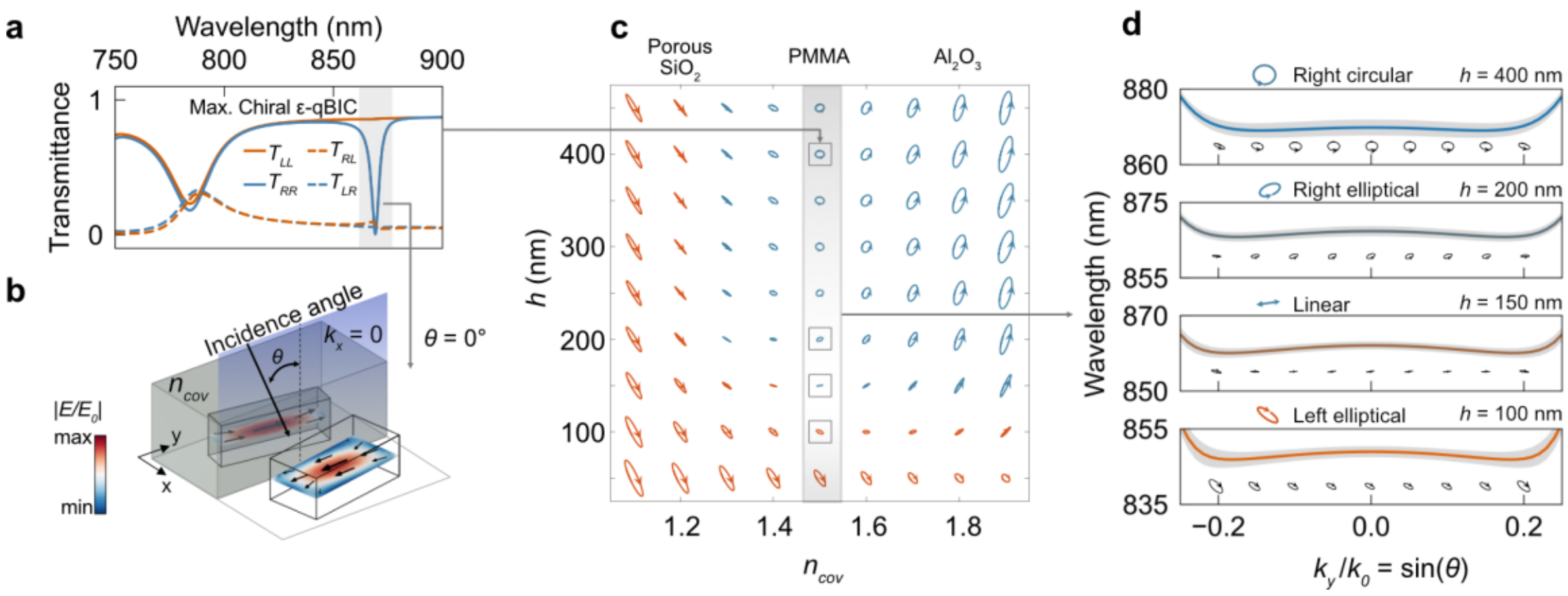

**Figure S2 Analysis of ε-qBIC polarization in metasurfaces with different claddings.** **a** Simulated transmittance spectra for a metasurface with PMMA cladding ($n$ = 1.5) and maximally chiral ε-qBIC, at the wavelength of 870 nm. **b** Schematic of the structure under illumination at an incidence angle $\theta$ in the $k_x$ = 0 plane, where the resonance exhibits a flatband in (d). The simulated electric near-field enhancement of the chiral ε-qBIC is shown in the middle cross-section of the nanorods for $\theta$ = 0°. **c** Simulated far-field polarization map of the right-handed metasurface as a function of the cladding thickness $h$ and refractive index $n_{cov}$. Representative cladding materials include porous $SiO_2$ ($n_{cov}$ = 1.2), PMMA ($n_{cov}$ = 1.5), and $Al_2O_3$ ($n_{cov}$ = 1.8). The shaded grey gradient indicates the evolution of the polarization state as the PMMA layer thickness varies. Blue ellipses correspond to the right-handed metasurface response, whereas red ellipses represent the opposite handedness polarization. The size of each ellipse is inversely proportional to the Q-factor. **d** Simulated $k_y$-dispersions for four polarization states with different $h$ reveal flatband behavior that persists up to $k_y/k_0 \approx 0.18$. The grey region denotes the full width at half maximum (FWHM) of the resonance. The corresponding polarization states are shown below each curve.

## 4 Reflectance of the ε-qBIC

At the ε-qBIC in the maximally chiral structure shown in Figure 1b, RCP light is reflected while preserving it handedness (Figure S3), which enables angle-resolved reflectance measurements.

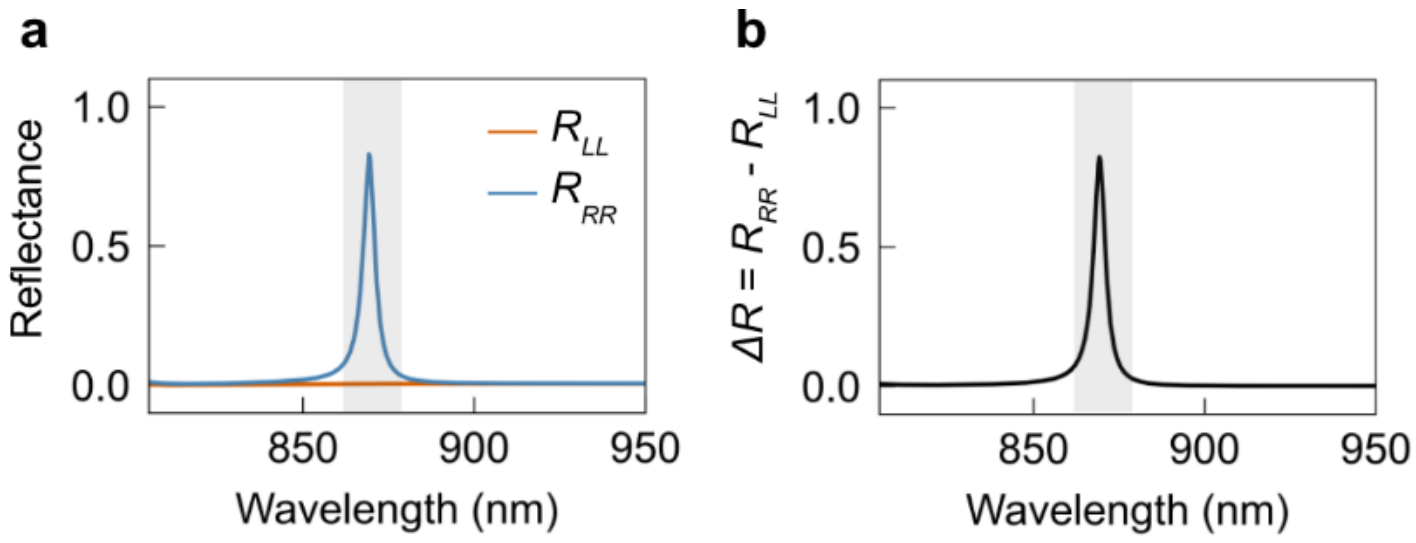

**Figure S3. Reflectance of the ε-qBIC.** Simulated (**a**) co-polarized reflectance and (**b**) reflectance difference $\Delta R = R_{RR} - R_{LL}$ for the maximally chiral ε-qBIC. The grey shaded region indicate an ε-qBIC at a wavelength of 870 nm.

## 5 Fabrication robustness of the metasurface with chiral ε-qBICs

Our platform exhibits enhanced robustness against fabrication variations with respect to the PMMA thickness. The maximal chiral response is maintained over a PMMA height range of 270–600 nm (Figure S4a). In comparison, the nanorod configuration with different heights maintains a maximal chiral response only over a narrow height range of 157–170 nm for the taller nanorod (Figure S4b) [10]. Therefore, our platform represents more than a 25-fold increase in fabrication tolerance. Importantly, the ε-qBIC remains spectrally stable at the target wavelength, well-isolated from nearby other modes, and retains a high Q-factor across the full PMMA height range (Figure S4c, d), whereas the chiral response in Ref [10] is perturbed by the mode coupling when the nanorod height is 200 nm (top set of curves in Figure S4e, f). Furthermore, as the PMMA height can be precisely controlled using the RIE process, our platform provides a flexible and precise method to achieve the desired chiral response. For left elliptical polarization state, our design also exhibits a larger difference in co-polarized transmittance and a higher Q-factor relative to the Ref[10]. These results highlight that our platform not only significantly relaxes fabrication constraints but also preserves strong, high-Q chiral optical responses, demonstrating its practical advantage for tunable and robust chiral metasurfaces.

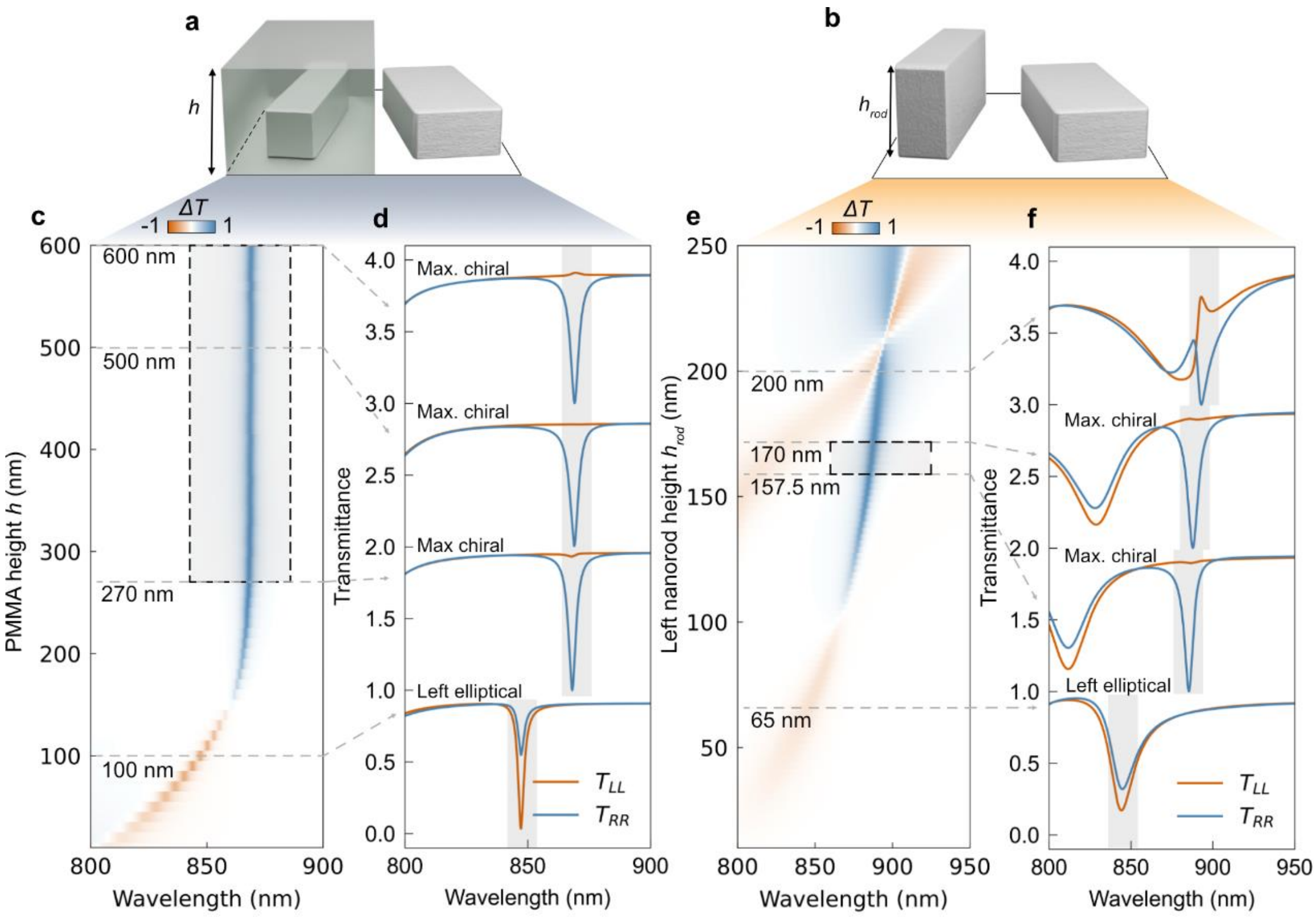

**Figure S4.** Comparison of (**a**) the ε-qBIC metasurface and (**b**) the chiral metasurface featuring nanorods with different heights [10]. Colormaps of the transmittance difference, $\Delta T = T_{LL} - T_{RR}$, as a function of wavelength and structural parameters: (**c**) PMMA thickness $h$, and (**e**) left nanorod height $h_{rod}$. (**d**) and (**f**) show the corresponding transmittance spectra. The grey-shaded regions in colormaps in (**c**) and (**e**) indicate the height range of maximum chiral response with $\Delta T \approx 1$, while the grey-shaded regions in (**d**) and (**f**) indicate the qBIC position. Structural parameters of the ε-qBIC structure: each nanorod has an identical length $L$ = 288 nm, and height $H$ = 130 nm, but distinct widths of $W_1$ = 65 nm and $W_2$ = 145 nm, with an opening angle $\alpha$ = 7.5°.

## 6 Effect of the opening angle between nanorods

To achieve different polarization states, the opening angle $\alpha$ between two nanorods of the ε-qBIC metasurface can be adjusted while keeping the PMMA height fixed at 400 nm (Figure S5). As the angle $\alpha$ is tuned from -7.5° to 7.5°, the polarization evolves from left circular, through left elliptical and linear, to right elliptical and right circular polarization. Right circular polarization corresponds to the maximum chirality discussed in the main text, while left circular polarization exhibits the opposite handedness. When changing the height of PMMA, the polarization can be further changed, offering an additional degree of freedom. This dual-control strategy highlights the versatility of our platform for precise and flexible manipulation of polarization states.

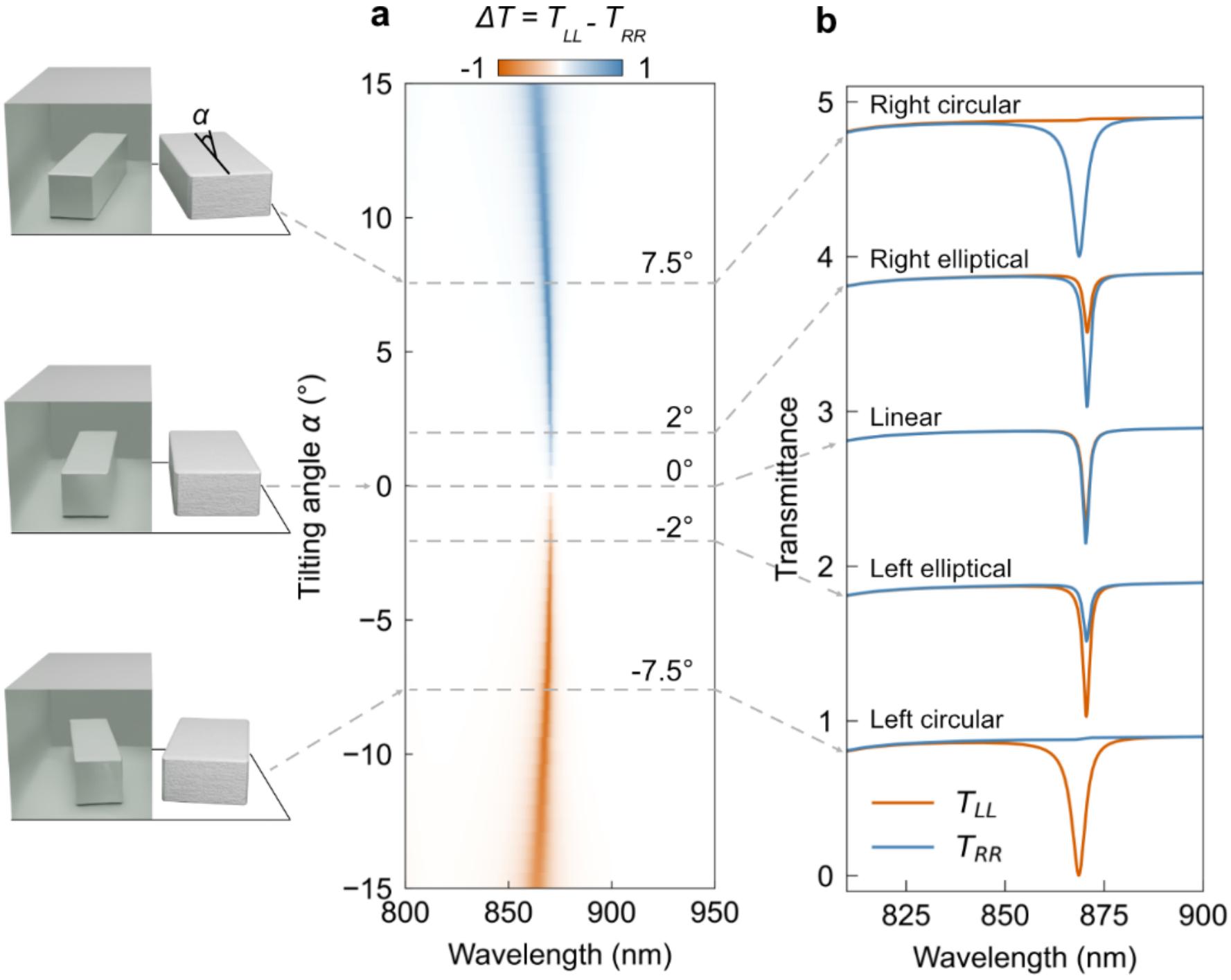

**Figure S5. Effect of opening angle $\alpha$ on the chiral ε-qBIC. a** Co-polarizated transmittance difference $\Delta T$ as a function of opening angle $\alpha$ and wavelength. **b** Corresponding simulated transmittance spectra $T_{LL}$ and $T_{RR}$ for different polarization states. Structural parameters of the ε-qBIC structure: each nanorod has an identical length $L$ = 288 nm, and height $H$ = 130 nm, but distinct widths of $W_1$ = 65 nm and $W_2$ = 145 nm; PMMA height $h$ = 400 nm.

## 7 Experimental co-polarized reflectance difference

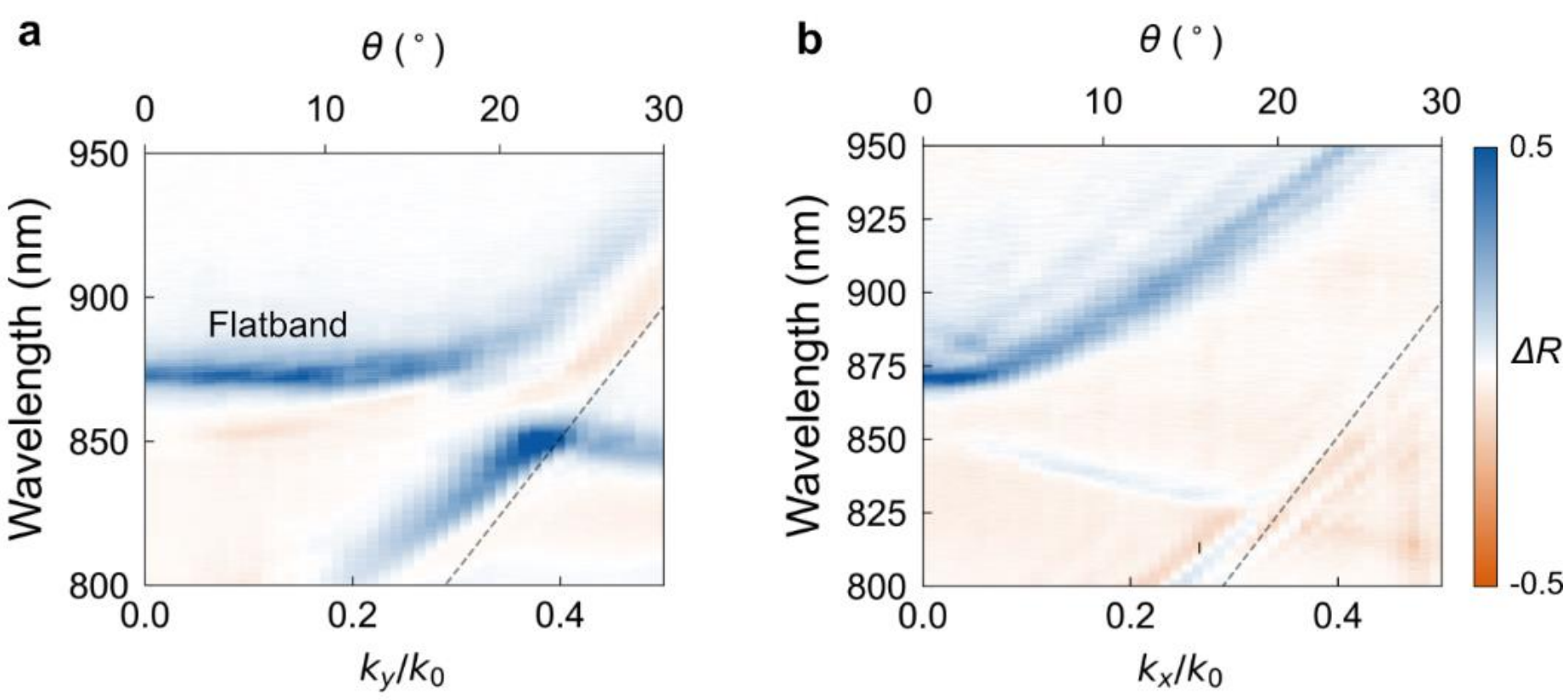

**Figure S6. Experimental co-polarized reflectance difference of the maximally chiral ε-qBIC**. The reflectance difference $\Delta R = R_{RR} - R_{LL}$ shows flatband dispersion along the $k_y/k_0$ direction ($k_x = 0$) in (**a**) and parabolic dispersion along the $k_x/k_0$ direction ($k_y = 0$) in (**b**).

## 8 Circularly polarized ε-qBICs with parabolic dispersion

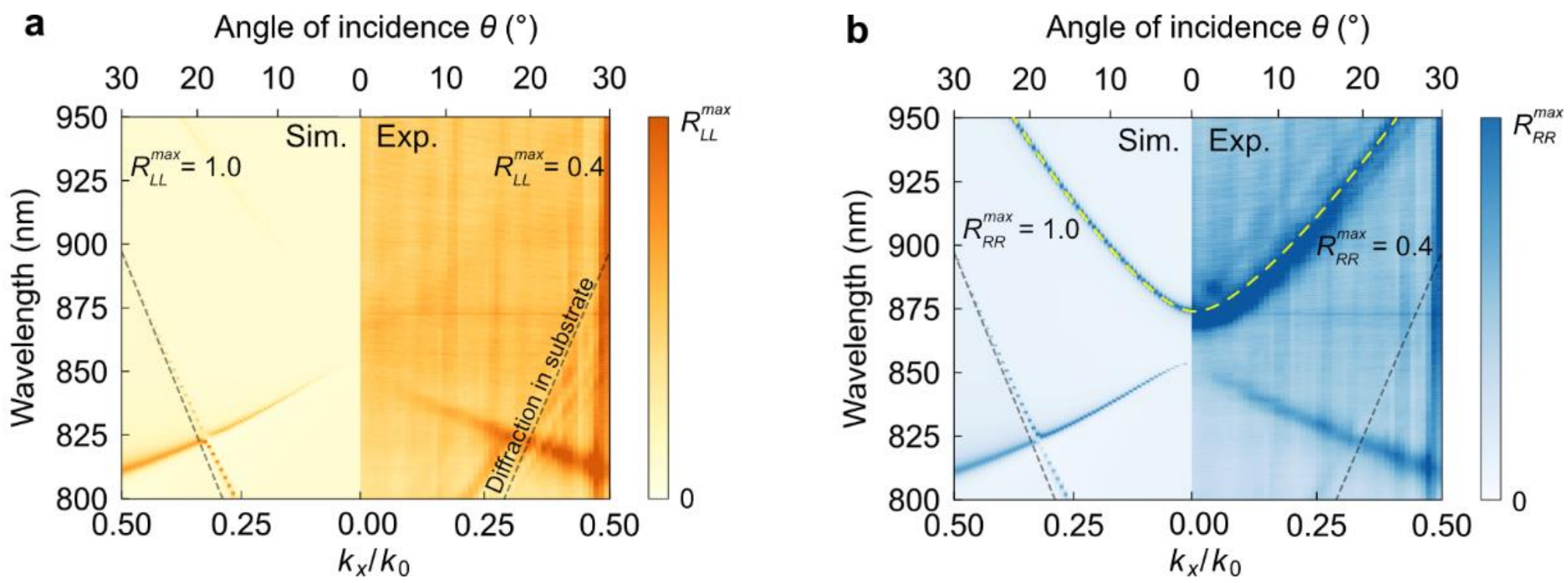

**Figure S7. Parabolic dispersion of the circularly polarized ε-qBIC.** Simulated and experimental co-polarized reflectance spectra $R_{LL}$ (**a**) and $R_{RR}$ (**b**) for obliquely incident light along the $k_x/k_0$ direction ($k_y = 0$). The dashed lines indicate the diffraction cutoff associated with the $SiO_2$ substrate. The yellow dashed line in (**b**) highlights the parabolic dispersion of the experimental chiral ε-qBIC.

## 9 Parabolic dispersion of the chiral ε-qBIC with different polarizations

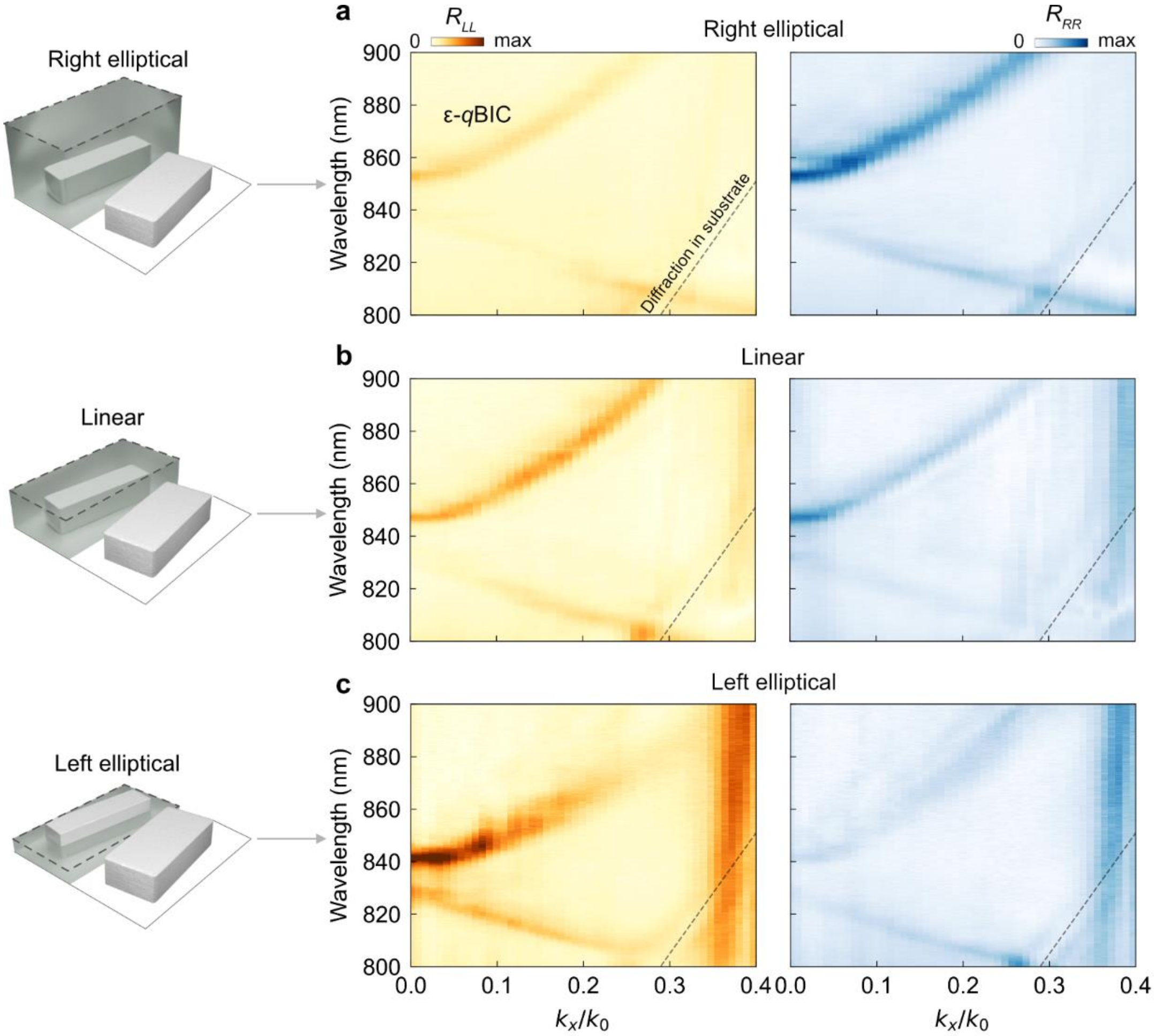

**Figure S8. Experimental parabolic dispersion of the chiral ε-qBIC with different PMMA thicknesses.** Co-polarized reflectance spectra $R_{LL}$ (left panel) and $R_{RR}$ (right panel) as functions of $k_x/k_0$ ($k_y = 0$), for the metasurface shown in Figure 4 of the main text. The ε-qBIC exhibits three different polarization states: right elliptical, linear, and left elliptical. The left insets show the corresponding unit cells with the PMMA cladding of different thicknesses.

## 10 Spatial chiral encoding under oblique incidence

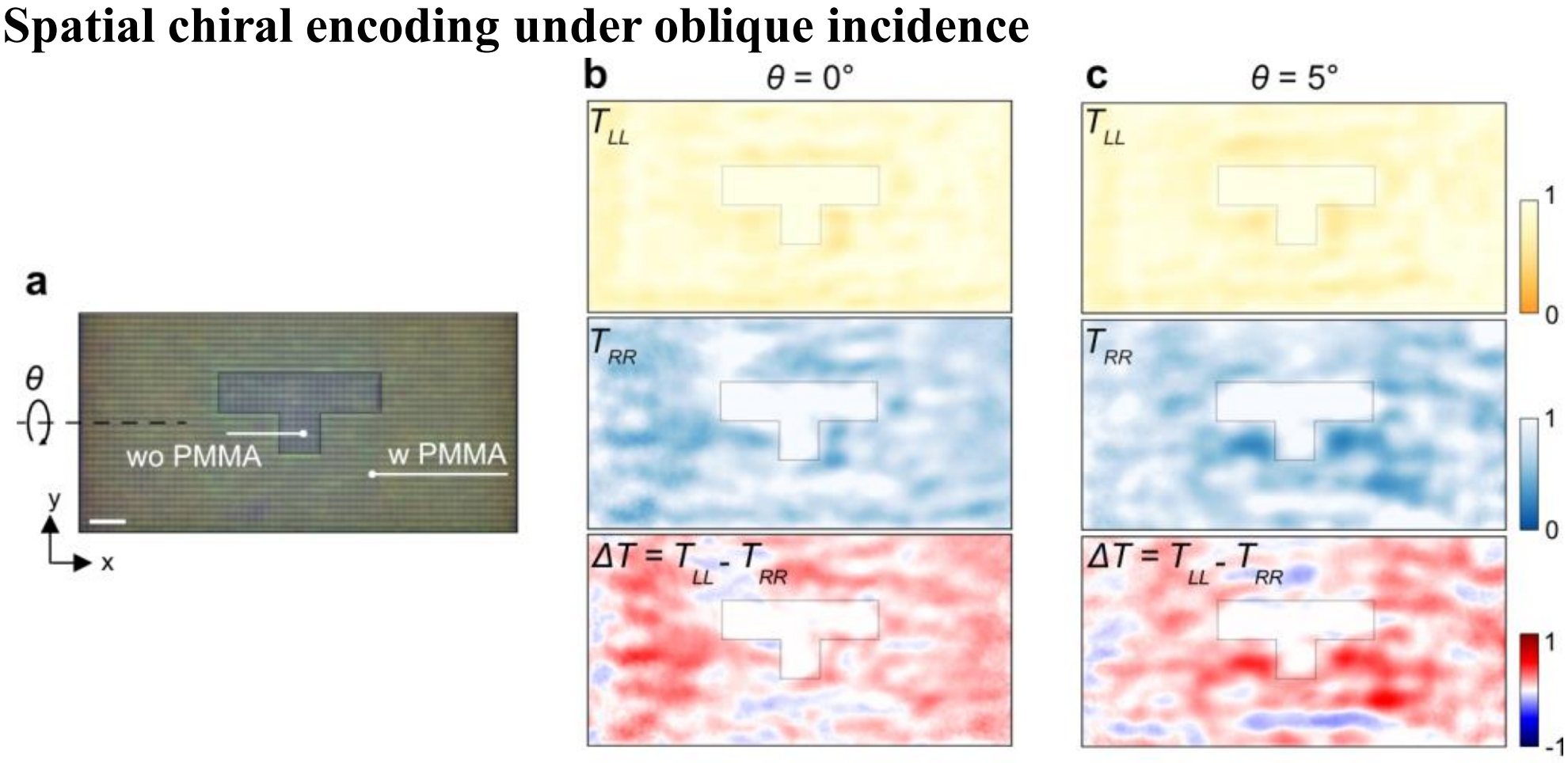

**Figure S9. Chiral imaging of the rotated ε-qBIC metasurfac. a** Schematic of the rotated metasurface rotated by an angle $\theta$ with the rotation axis along the middle line of the metasurface. **b** Chiral imaging based on co-polarized transmittance $T_{RR}$, $T_{LL}$, and their difference $\Delta T$ for $\theta = 0°$ (**b**) and $\theta = 5°$ (**c**), respectively at $\lambda_0 = 860$ nm. The solid lines delineate the T-shaped region. For the $\Delta T$ image, surrounding PMMA-covered area predominantly appears red ($\Delta T > 0$) for both rotation angles, indicating that the chirality remains robust under oblique incidence.

## 11 Chiral vibrational strong coupling

Given the ability of PMMA to incorporate various molecules, we investigate our platform for potential chiral strong coupling. The new metasurface has the following dimensions: $L$=245 nm, $W_1$=125 nm, $W_2$ = 60 nm, $H$ = 110 nm, $\Lambda_x = \Lambda_y$ = 455 nm, $\alpha$ = 8°, and $h$ = 315 nm. As shown in Figure S10b, the ε-qBIC preserves its circular polarization in a wide $k$-space range. Thus, our system has the advantage of the angular robust far-field polarizations in $k$-space. The characteristic half-pipe-dispersion of the mode wavelength allows the chiral ε-qBIC to be tuned from 780 nm to 810 nm via oblique incidence along the $k_x/k_0$ ($k_y$ = 0) direction (red shaded area in Figures S10b-d), while maintaining a high and stable Q-factor (Figure S10d). This enables continuous tuning of the ε-qBIC to overlap with material features, such as excitonic or molecular absorption lines, by adjusting the incidence angle, thereby facilitating chiral light-matter interaction.

To illustrate the chiral vibrational strong coupling, we choose a Copper Phthalocyanine (CuPc) molecule[11–13] having the resonance around 788 nm (Figure S10e). The molecules are assumed to be uniformly dispersed within the PMMA layer prior to spin-coating. The optical response of CuPc molecule is modeled using a classical Lorentz oscillator[14-15]:

$$\varepsilon_{CuPc} = \varepsilon_{\infty} + \frac{N_{CuPc} f_0 \omega_0^2}{\omega^2 - \omega_0^2 - i\gamma_0 \omega}$$

where $\varepsilon_{\infty}$ is the background relative permittivity of CuPc molecule of 2.15, $f_0$ is the strength coefficient of 0.0165, $\omega_0$ is the Lorentz resonance frequency of $2.39 \times 10^{15}$ rad/s, $\gamma_0$ is the Lorentz damping rate of $4.41 \times 10^{13}$ rad/s, and $N_{CuPc}$ is the relative concentration coefficient of 0.2, 0.4 and 0.6, representing three different kinds of concentration (Figure S10f). All parameters are selected to closely align with the experimental data.

The spectral tunability of the right-handed chiral ε-qBIC metasurface is implemented through oblique incidence across the CuPc vibrational resonance. Figure S10g shows circular dichroism $CD = (T_{LL} - T_{RR})/(T_{LL} + T_{RR})$ for different relative concentrations of CuPc. The results indicate that coupling between the photonic ε-qBIC and the molecular vibrational resonance ($\lambda_{CuPc}$ = 788 nm) leads to the formation of upper and lower chiral polariton branches. When the relative concentration coefficient $N_{CuPc}$ is 0, the chirality of ε-qBIC remains almost unaffected up to an incidence angle of 11.5°, corresponding to $k_x/k_0$ = 0.2. The Rabi splitting exhibits a progressive enhancement as the relative molecular concentration changes from 0.2 to 0.4 and subsequently to 0.6.

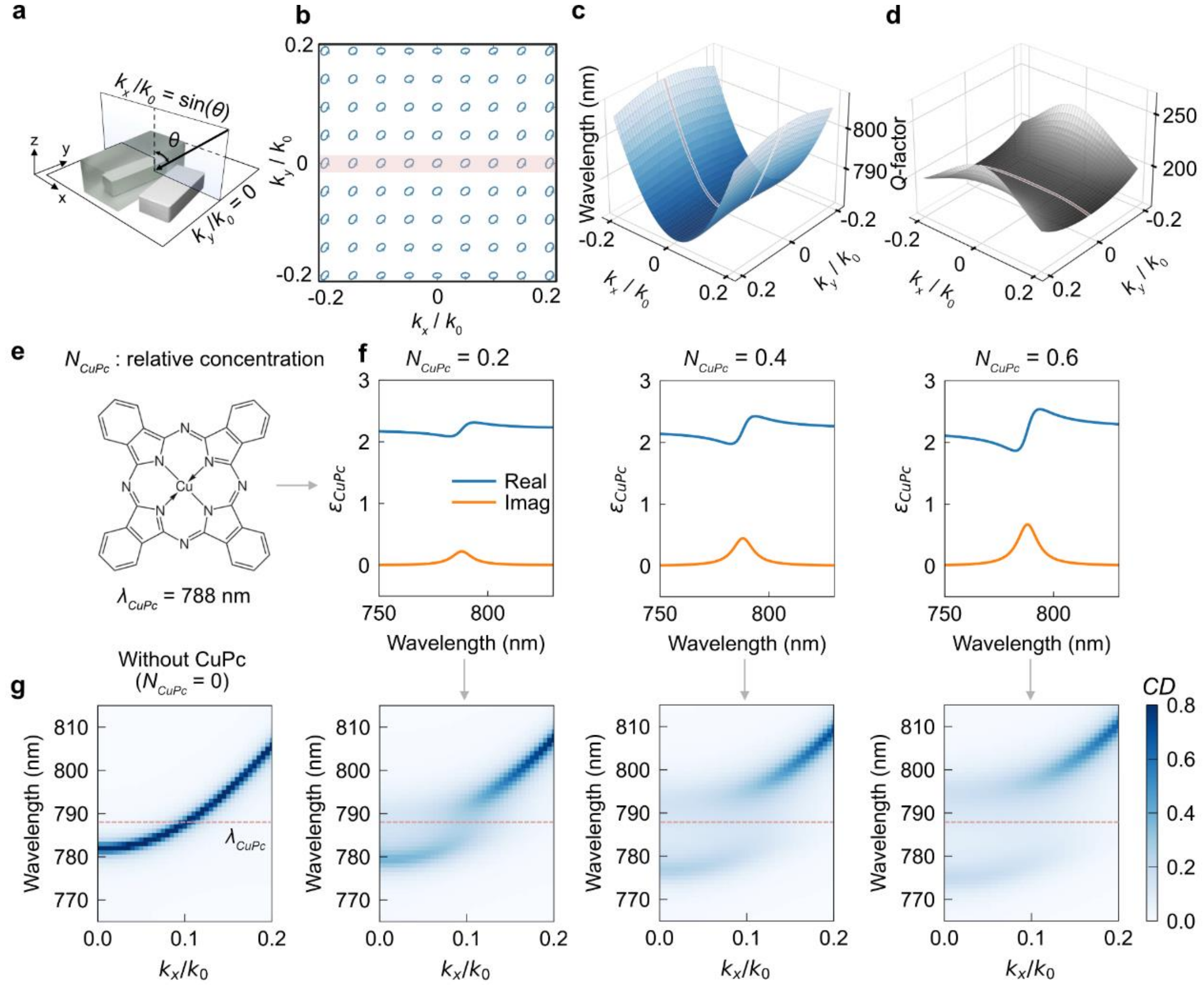

**Figure S10. Simulated chiral vibrational strong coupling.** **a** Schematic of maximally chiral structure for oblique incident light in $k_x$ direction ($k_y$ = 0). Simulated far-field polarization (**b**), resonance position (**c**), and Q-factor (**d**) in $k$-space for the maximally chiral ε-qBIC. The red shaded area highlights the parabolic region. **e** The structure of CuPc molecule with a vibrational resonance at 788 nm. **f** Permittivity of the CuPc molecule for three relative concentrations ($N_{CuPc}$ = 0.2, 0.4 and 0.6) embedded in the PMMA cladding. **g** Simulated angle-resolved CD as a function of $k_x/k_0$ for corresponding concentrations ($N_{CuPc}$ = 0, 0.2, 0.4 and 0.6).

## 12 Measurement setup for normal-incidence spectroscopy

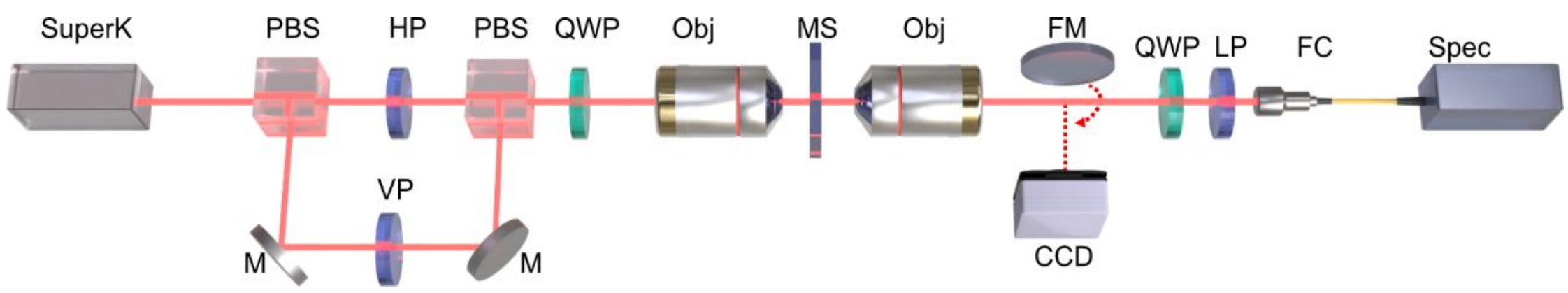

**Figure S11. Schematic of optical characterization.** SuperK: supercontinuum white light laser, PBS: polarizing beam splitter, HP: horizontal linear polarizer, VP: vertical linear polarizer, M: mirror, MS: metasurface, Obj: objective, FM: flip mirror, QWP: quarter wave plate, FC: fiber coupler, CCD: charge-coupled device, Spec: spectrometer. RCP or LCP light is produced by blocking either the HP or VP beam path.

## 13 Measurement setup for *k*-space spectroscopy

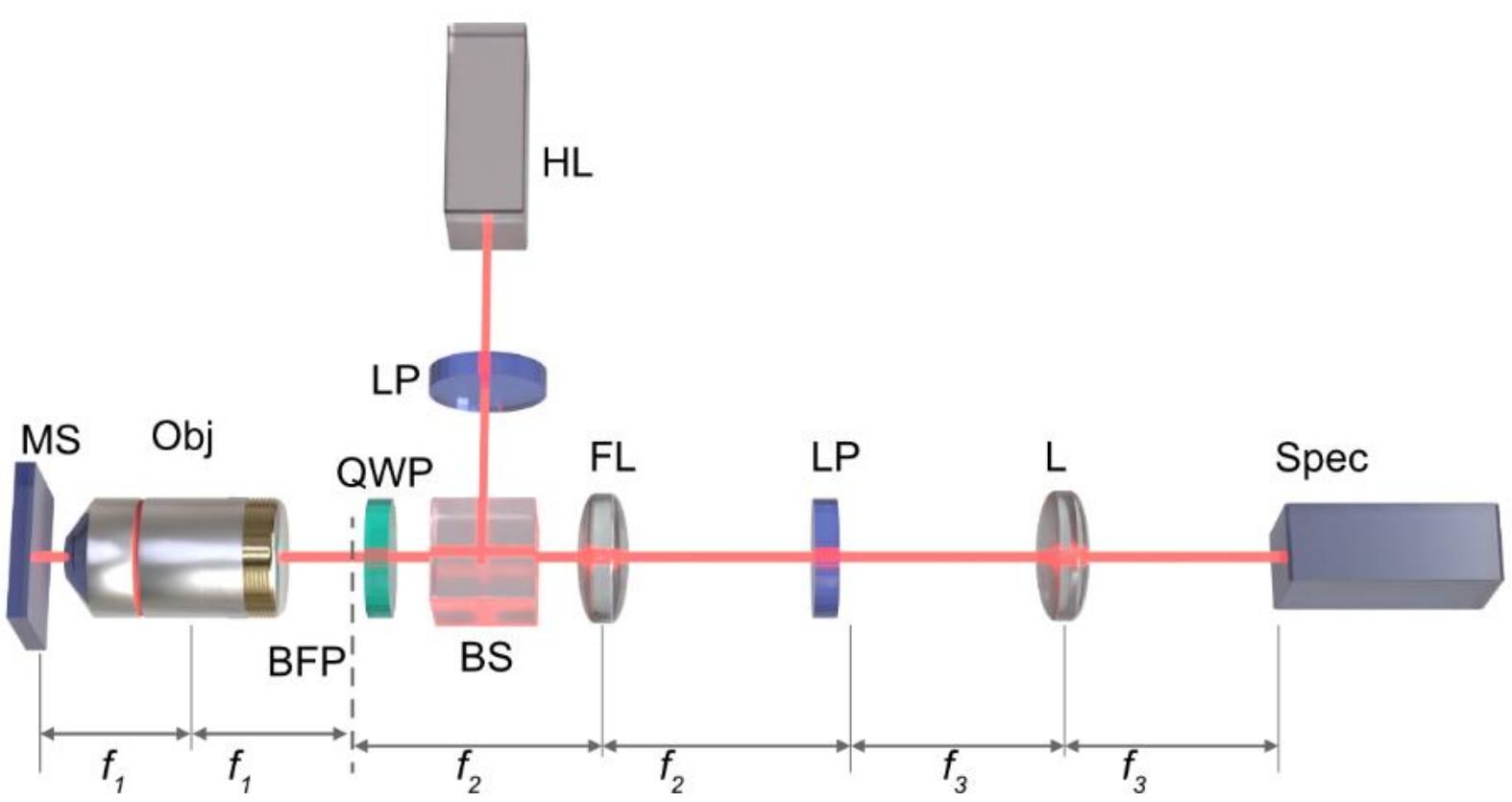

**Figure S12. Schematic of the angle-resolved reflection measurement setup.** MS: metasurface, Obj: 60× objective (NA = 0.95), BFP: back focal plane, QWP: quarter-wave plate, BS: 50:50 beam splitter, HL: halogen lamp; LP: linear polarizer, FL: Fourier lens, L: lens, Spec: spectrometer. Collimated white light from the HL is converted to circular polarization by the LP and QWP, directed onto the sample through the BS and Obj, and the reflected light is collected by the same Obj. The signal passes through the QWP, BS, and analyzer LP for polarization analysis. The back focal plane is imaged by the FL and relayed to the spectrometer via a 4f system.